\documentclass[a4paper,fleqn]{cas-dc}

\usepackage[authoryear]{natbib}
\usepackage{bm}

\def\tsc#1{\csdef{#1}{\textsc{\lowercase{#1}}\xspace}}
\tsc{WGM}
\tsc{QE}

\begin{document}
\let\WriteBookmarks\relax
\def\floatpagepagefraction{1}
\def\textpagefraction{.001}

\shorttitle{Assessing the enrichment of heat-producing elements in deep-mantle structures using a directional geoneutrino detector}

\shortauthors{Z. Xu et al}  

\title [mode = title]{Assessing the enrichment of heat-producing elements in deep-mantle structures using a directional geoneutrino detector}  

%

\author[1]{Zhihao Xu}[orcid=0009-0008-6580-0259]
\cormark[1]
\ead{xu.zhihao.t1@dc.tohoku.ac.jp}
\credit{Conceptualization, Investigation, Methodology, Data curation, Formal analysis, Visualization, Software, Writing – original draft, Writing – review \& editing, Funding acquisition}

\author[1]{Misaki Hosoya}[orcid=0009-0002-4421-5911]
\credit{Investigation, Methodology, Data curation, Validation, Visualization, Software, Writing – review \& editing}

\author[2, 3, 4]{William F. McDonough}[orcid=0000-0001-9154-3673]
\credit{Conceptualization, Methodology, Validation, Supervision, Writing – review \& editing, Funding acquisition}

\author[5]{Taichi Sakai}[orcid=0000-0003-2855-6505]
\credit{Investigation, Software, Data curation, Writing – review \& editing}

\author[1]{Hiroko Watanabe}[orcid=0000-0002-2363-5637]
\credit{Conceptualization, Methodology, Validation, Supervision, Writing – review \& editing, Funding acquisition}

\affiliation[1]{organization={Research Center for Neutrino Science, Tohoku University},
            city={Sendai},
            postcode={980-8578}, 
            state={Miyagi},
            country={Japan}}

\affiliation[2]{organization={Center for Geoneutrino Research, Institute of Oceanology, Chinese Academy of Sciences},
            city={Qingdao},
            postcode={266237}, 
            state={Shandong},
            country={China}}
        
\affiliation[3]{organization={Department of Earth Science, Tohoku University},
            city={Sendai},
            postcode={980-0578}, 
            state={Miyagi},
            country={Japan}}

\affiliation[4]{organization={Department of Geological, Environmental, and Planetary Sciences, University of Maryland, College Park},
            city={College Park},
            postcode={20742}, 
            state={MD},
            country={USA}}
        
\affiliation[5]{organization={High Energy Accelerator Research Organization (KEK)},
            city={Tsukuba},
            postcode={305-0801}, 
            state={Ibaraki},
            country={Japan}}



\begin{abstract}
Determining the spatial distribution of heat-producing elements (HPEs) within the Earth is critical for understanding the planet's thermal and chemical evolution.
A central debate is whether domains in the deep mantle, particularly the Large Low-Velocity Provinces (LLVP), represent anomalous, radiogenically enriched reservoirs.
Variation in the surface flux of geoneutrinos provide a direct probe of Earth's internal radioactivity.
However, current continental located detectors measure only the angle-integrated flux, leading to strong spatial trade-offs that substantially limit the ability of practical networks to uniquely resolve mantle compositional heterogeneity.
In this study, we quantify the potential of directional geoneutrino detection to overcome this limitation.
By evaluating realistic LLVP geometries in the context of the proposed Ocean Bottom Detector (OBD), we demonstrate that resolving the incoming direction of geoneutrinos can substantially improve the precision with which HPE enrichment in LLVPs can be constrained relative to the surrounding mantle, for a range of target angular resolutions envisioned for a future directional geoneutrino detector.
These results indicate that future directional geoneutrino measurements could help determine whether LLVP host enhanced HPE abundances and assess their contribution to Earth’s radiogenic heat budget.
Such measurements would provide a new observational constraint on the chemical heterogeneity of the deep mantle and its role in Earth's long-term thermal evolution.
\end{abstract}




\maketitle

\newcommand{\add}[1]{\textcolor{ForestGreen}{#1}}
\newcommand{\del}[1]{\textcolor{red}{\sout{#1}}}

\def\tsc#1{\csdef{#1}{\textsc{\lowercase{#1}}\xspace}}
\tsc{WGM}
\tsc{QE}


\section{Introduction}
\label{sec:intro}
The radioactive decay of heat-producing elements (HPEs; predominantly U, Th, and K) within the Earth supplies roughly 40\% of Earth's present-day internal heat budget that drives the geodynamo, mantle convection and plate tectonics \citep{mcdonough2026earth}.
A fundamental challenge in geoscientific research is determining the spatial distribution of these elements.
Whether the mantle is thoroughly mixed by billions of years of convection or still preserves chemically distinct reservoirs at depth remains a central question, as this controls our understanding of Earth’s thermal and compositional evolution.
A key focus of this debate is the large-scale heterogeneity observed at the base of the mantle.
The Large Low-Velocity Provinces (LLVP), two continent-scale regions of anomalously low seismic velocity in the lowermost mantle, with one centered beneath the Pacific Ocean and the other beneath the African continent, are the most prominent candidates for such deep thermochemical reservoirs \citep{garnero2016continent}.
Identifying whether these domains are enriched in HPEs is essential to reconciling seismic observations with Earth's bulk composition and heat flow, for which geoneutrino detection has been highlighted as a potential approach \citep{ferrick2023defining}.

Geoneutrinos ($\bar{\nu}_e$) emitted during beta decay chains offer the only direct, unattenuated observational probe of this radiogenic power \citep{eder1966terrestrial}.
Since the first observation of geoneutrinos by the KamLAND experiment in 2005 \citep{araki2005experimental}, kiloton-scale liquid-scintillator neutrino detectors have constituted the exclusive technology established for this field.
While completed and ongoing experiments, including KamLAND \citep{kamland2022geonu}, Borexino \citep{borexino2020geonu}, SNO+ \citep{sno2026geonu}, and JUNO \citep{juno2025geonu}, provide essential constraints on the global geoneutrino flux, they cannot resolve the spatial distribution of the signal.
These experiments detect geoneutrinos via the inverse beta decay (IBD) reaction:
\begin{equation}
    \bar{\nu}_e + p \rightarrow e^+ + n,
    \label{ibd}
\end{equation}
where the incoming direction of neutrino is encoded in the kinematic displacement between the prompt positron ($e^+$) signal and the delayed neutron ($n$) capture \citep{vogel1999angular}.
However, because multiple neutron scatterings and finite detector spatial resolution heavily blur this tiny displacement vector, these instruments measure only the angle-integrated geoneutrino flux, acting purely as rate-only detectors \citep{tanaka20146li}.
This limitation results in an inherent non-uniqueness in the interpretation of geoneutrino data, as the same event rate can be produced by either a chemically homogeneous mantle or a heterogeneous one where depleted regions are compensated by HPE-enriched domains \citep{roskovec2018testing}.
Although deploying rate-only detectors across multiple locations can help constrain the overall radiogenic heat of the bulk mantle, localized HPE enrichment in mantle structures is difficult to isolate \citep{sramek2013geophysical}.

Directional geoneutrino detection introduces a new observable capable of overcoming the directionally blind nature of current detectors, and several emerging experimental concepts have been proposed to achieve this angular resolution \citep{tanaka20146li,safdi2015directional,leyton2017exploring,wang2020hunting,duvall2024directional,yepez2026algorithm,crow2026enhancing,sun2026potassium}.
While previous studies successfully demonstrated the feasibility of separating mantle signals from crustal backgrounds using angular resolution \citep{fields2006imaging,shimizu2007directional,leyton2017exploring,sun2026potassium}, these efforts primarily focused on simplified homogeneous or 1D layered mantle, lacking lateral variations.
An initial proof-of-concept study recently suggested that directional detection could isolate LLVP signals from the ambient mantle \citep{xu2026proceedings}.
However, this previous study relied on idealized cylinder-like LLVP geometries, considered a fixed HPE inventory, and did not account for non-geoneutrino backgrounds or uncertainties in the Earth model.
These simplifications may have limited the robustness and generality of the sensitivity estimates.

In this work, we quantify the potential of directional geoneutrino detection to constrain HPE enrichment in deep-mantle structures using tomographically derived LLVP geometries, while accounting for uncertainties in the Earth model and contributions from non-geoneutrino backgrounds.
Chap. \ref{sec:modeling} describes the construction of Earth models, including a homogeneous mantle model and a heterogeneous mantle model incorporating LLVPs, which serve as the reference models for the subsequent sensitivity studies.
Chap. \ref{sec:methods} describes the calculation of geoneutrino fluxes and formulates the inverse problem, including the free parameters to be constrained.
Chap. \ref{sec:detector} then describes the experimental framework considered in this study, including the detector configuration and angular resolution.
Finally, Chap. \ref{sec:directional_solution} evaluates the sensitivity of directional geoneutrino detection to HPE enrichment in the LLVPs, quantifying the precision with which the HPE abundances of LLVPs and the ambient mantle can be determined and sensitivity with which a homogeneous mantle model can be excluded.

\section{Earth Modeling}
\label{sec:modeling}

\subsection{Basic Structure and LLVP Geometry}
To establish a baseline for geoneutrino flux calculations and provide a reference model for comparison, we first construct a basic Earth model with a chemically homogeneous mantle.
Specifically, we adopt the Preliminary Reference Earth Model (PREM) \citep{dziewonski1981prem} to define the density profile of the mantle, while using the Earth Crustal Model 1 (ECM1) \citep{mooney2023ecm} to construct the crustal structure and density distribution.
To account for flux uncertainties introduced by crustal geometry, we assign a 12\% uncertainty to the thickness of each crustal layer following \citet{huang2013reference}.

Due to the 1.806 MeV energy threshold of the IBD reaction \eqref{ibd}, only geoneutrinos from $^{238}\text{U}$ and $^{232}\text{Th}$ decay chains are detectable, while those from $^{40}\text{K}$ and other radioisotopes fall below this limit \citep{raghavan1998measuring}.
Since both U and Th are strongly lithophile, they are effectively partitioned into the silicate Earth rather than the metallic core \citep{mcdonough2003compositional,wipperfurth2018earth}.
Therefore, geoneutrinos from the core are negligible, and the core structure is therefore omitted from this study.

\begin{figure*}[ht]
    \centering
    \includegraphics[width=1\textwidth]{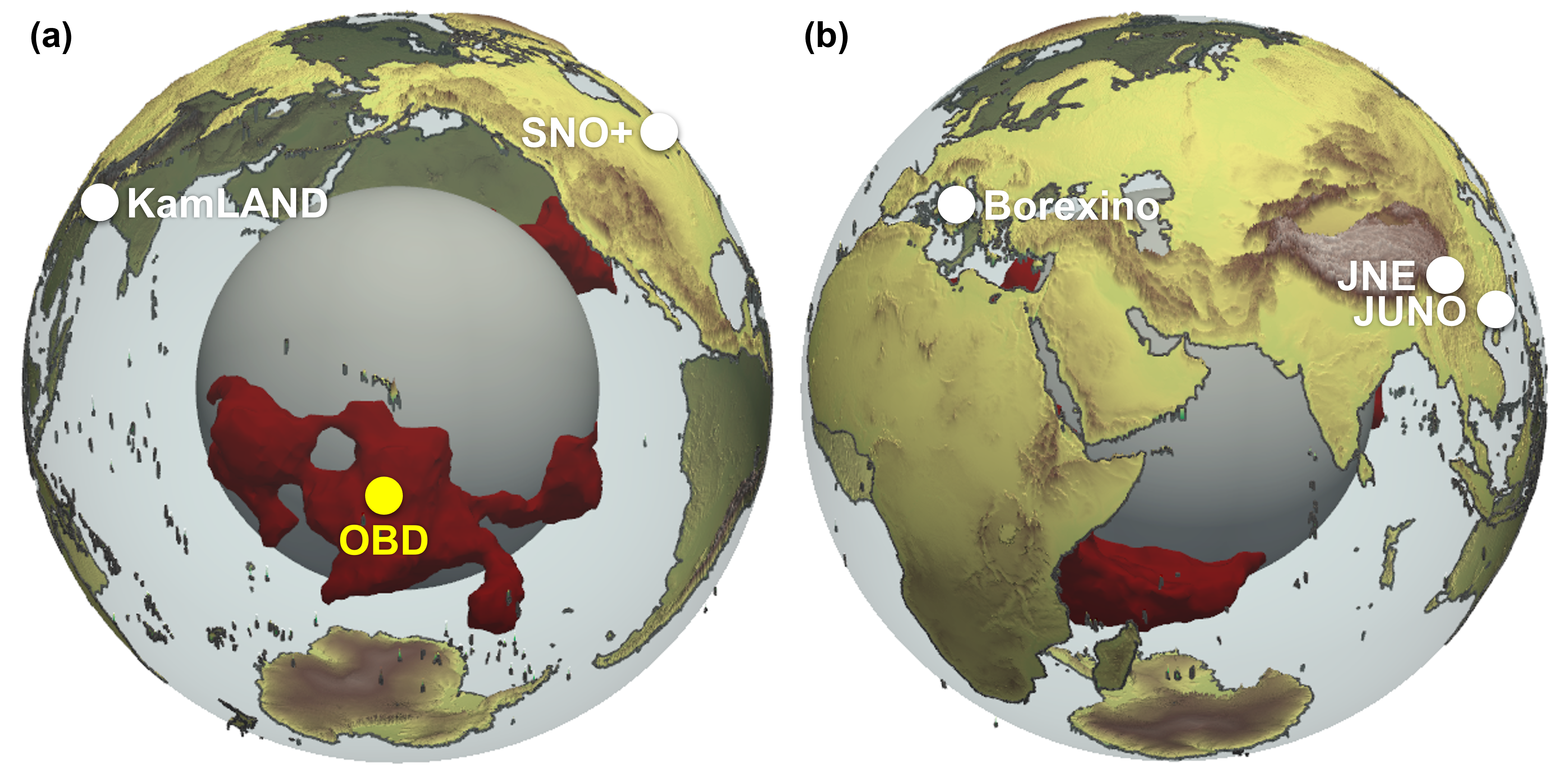}
    \caption{Spatial distribution of geoneutrino detectors relative to the LLVP domains in the lower mantle.
    Panels (a) and (b) present the Pacific-centric and Eastern Hemisphere-centric views, respectively.
    Red regions delineate the LLVP geometries defined by the tomographic consensus model.
    The yellow circle denotes the proposed Ocean Bottom Detector (OBD), and white circles represent decommissioned, existing, and under-construction continental geoneutrino detectors, including KamLAND, Borexino, SNO+, JUNO, and JNE.}
    \label{fig:U_distribution}
\end{figure*}

To investigate the sensitivity of geoneutrino signals to the HPE enrichment in LLVPs, we compare the geochemically homogeneous mantle reference case against 
a heterogeneous scenario constructed using seismic tomography vote maps \citep{cottaar2016morphology}.
Rather than applying a single seismic velocity threshold to a single tomography model, this vote map delineates LLVP geometries using consensus vote counts derived from five tomographic models: S40RTS \citep{ritsema2011s40rts}, Savani \citep{auer2014savani}, SEMUCBWM1 \citep{french2014whole}, SPani \citep{tesoniero2015hydration}, and ME2016-S \citep{moulik2016relationships}.
In this study, we extract LLVP boundaries at depths exceeding 1000 km using vote-count thresholds. 
Following \citet{cottaar2016morphology}, we adopt their definitions of the "consensus model" based on full agreement across all 5 tomographic models ($= 5$ votes). 
To prevent small, isolated velocity anomalies from cluttering the geometry, we apply a spatial continuity criterion to isolate the two largest contiguous structures, corresponding to the African LLVP and Pacific LLVP.
The spatial distribution of extracted LLVP geometry is shown in Fig.~\ref{fig:U_distribution}.

\subsection{HPE Inventory}
Given that the true spatial distribution of HPEs in the Earth remains largely unconstrained, we adopt an idealized mass-balance approach to construct test Earth models for our sensitivity evaluation.
Multiple independent studies have estimated the Bulk Silicate Earth (BSE) composition, including the estimates of \citet{mcdonough1995composition} and \citet{palme2014}, the updated analysis by \citet{mcdonough2026earth}, independent bootstrap analyses by \citet{lyubetskaya2007a,lyubetskaya2007b}, and an analysis allowing for a potentially hidden layer in the Earth's interior \citep{ferrick2023defining}.
These studies provide estimates of the present-day planetary radiogenic power that are broadly centered around $\sim 20$ TW, which we adopt as a reference constraint in our study.
For the purpose of the present benchmark study, our objective is not to establish a preferred BSE model or to quantify the full geochemical uncertainty across all existing literature estimates.
To this end, we adopt the updated BSE baseline model of \citet{mcdonough2026earth}, using a U abundance of $20.6 \pm 3.1$ ppb and a Th/U mass ratio of $3.78^{+0.12}_{-0.08}$ \citep{wipperfurth2018earth}, corresponding to a baseline planetary radiogenic power of $\sim 20$ TW.
We emphasize that the uncertainty associated with this baseline model is used to test sensitivity to local enrichment, rather than to encompass the complete diversity of proposed BSE compositions in the literature.

For the crustal component, we separate the contributions from the continental crust, oceanic crust, and overlying sediments.
For the continental crust, we adopt the global estimates from \cite{rudnick2014}, which explicitly differentiate the HPE abundances across the upper crust, middle crust, and lower crust.
The composition of the oceanic crust is prescribed using the values from \cite{white2014}.
Additionally, for the sediment layer, we employ the Global Subducting Sediments II (GLOSS-II) model \citep{plank2014} to represent the global average.
The specific values are summarized in Table \ref{tab:hpe_abundances}.

\begin{table*}[htb]
    \centering
    \caption{HPE abundances and reservoir masses adopted for the Earth models used in this study. Abundances for the crustal reservoirs are used in both models, while mantle abundances distinguish the homogeneous and heterogeneous models. Mass values for the crustal layers are based on ECM1 \citep{mooney2023ecm} assuming a 12\% uncertainty in thickness for each layer, and mantle masses are derived from PREM \citep{dziewonski1981prem}. References for HPE abundances are listed in the final column.}
    \label{tab:hpe_abundances}
    \renewcommand{\arraystretch}{1.2}
    \begin{tabular}{lcccc}
        \toprule
        Reservoir & U [ppm] & Th [ppm] & Mass [$10^{21}$ kg] & Abundance Ref. \\
        \midrule
        \multicolumn{5}{l}{\textbf{Crust}} \\
        \textit{Sediment} & $1.73\pm 0.09$ & $8.10\pm 0.59$ & $1.15^{+0.46}_{-0.26}$ & \citet{plank2014} \\
        \textit{Continental crust} & & & & \\
        \quad Upper crust & $2.7 \pm 0.6$ & $10.5 \pm 1.0$ & $6.19^{+0.79}_{-0.74}$ & \citet{rudnick2014} \\
        \quad Middle crust & $1.3 \pm 0.4$ & $6.5 \pm 0.5$ & $6.40^{+0.81}_{-0.76}$ & \citet{rudnick2014} \\
        \quad Lower crust & $0.2\pm 0.1$ & $1.2\pm 0.4$ & $6.77^{+0.86}_{-0.81}$ & \citet{rudnick2014} \\
        \textit{Oceanic crust} & $0.07\pm 0.02$ & $0.21\pm 0.06$ & $6.28^{+0.80}_{-0.75}$ & \citet{white2014} \\
        \midrule
        \multicolumn{5}{l}{\textbf{Mantle}} \\
        \textit{Homogeneous Model} & & & & \\
        \quad Bulk mantle & $0.014 \pm 0.003$ & $0.048^{+0.012}_{-0.011}$ & $4.01 \times 10^3$ & Mass-balance framework \\
        \addlinespace
        \textit{Heterogeneous Model} & & & & \\
        \quad Ambient mantle & $0.008 \pm 0.001$ & $0.023 \pm 0.006$ & $3.89 \times 10^3$ & \citet{arevalo2013simplified} \\
        \quad LLVP domain & $0.203^{+0.111}_{-0.104}$ & $0.836^{+0.445}_{-0.404}$ & $1.23 \times 10^2$ & Mass-balance framework  \\
        \bottomrule
    \end{tabular}
\end{table*}

The total mantle HPE inventory is then obtained by subtracting the total crustal HPE inventory from the BSE inventory.
This mantle HPE inventory is subsequently partitioned into two domains.
The ambient mantle outside LLVPs is assigned a MORB-source-like composition, based on the global MORB composition data of \citet{arevalo2010chemical} assuming 10\% partial melting \citep{arevalo2013simplified}, while the remaining HPEs are sequestered into the LLVP domains.
This mass-balance framework ensures that any observed variations in the geoneutrino signal arise solely from the spatial redistribution of HPEs, rather than changes in their total abundance.
Note that the HPE concentrations of MORB-source used here are higher than those estimated by \citet{salters2004composition} and \citet{workman2005major}.
This yields relatively lower HPE concentrations in the LLVPs, providing a relatively conservative assessment of their detectability.

\begin{figure*}[!htbp]
    \centering
    \includegraphics[width=\textwidth]{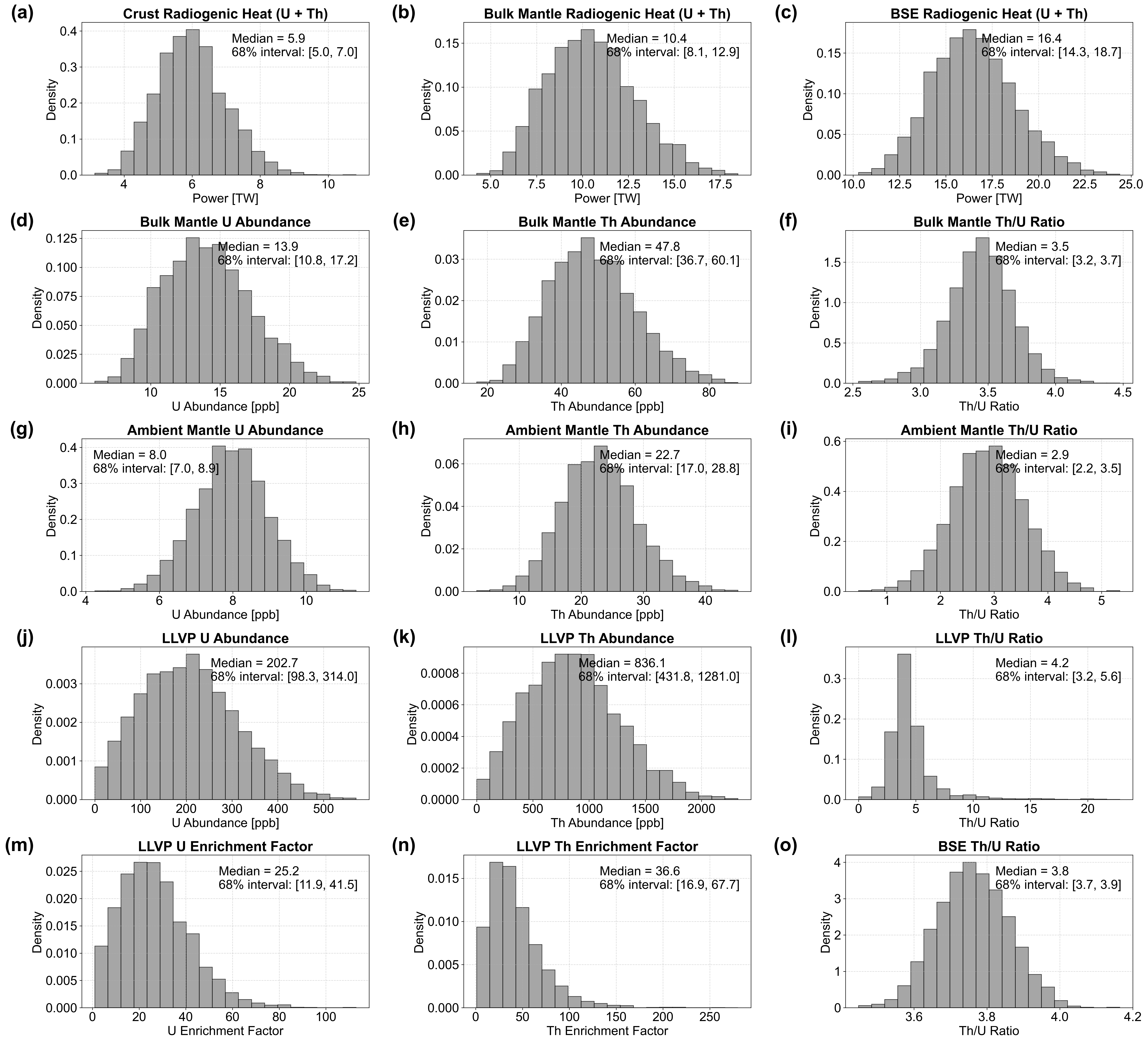}
    \caption{Probability density distributions of the heat-producing element (HPE) inventory, radiogenic power, and partitioning parameters obtained from the Monte Carlo sampling. 
    (a--c) Radiogenic heat generated by U and Th in the crust, the bulk mantle, and the bulk silicate Earth (BSE), respectively. The contribution from K is not included because its neutrinos are undetectable with current technologies.
    (d--f) U and Th abundances, and the corresponding Th/U ratio in the bulk mantle. 
    (g--i) U and Th abundances, and the corresponding Th/U ratio in the ambient mantle. 
    (j--l) U and Th abundances, and the corresponding Th/U ratio sequestered within the LLVP domains. 
    (m--n) Enrichment factors of U ($E^\mathrm{U}$) and Th ($E^\mathrm{Th}$) in the LLVPs relative to the ambient mantle. 
    (o) Th/U ratio of the BSE.}
    \label{fig:pdfs}
\end{figure*}

To explicitly account for the uncertainties in HPE concentrations, we employ a Monte Carlo sampling approach.
A large ensemble of input parameter sets is generated by drawing samples from probability distributions defined by the reported median values and associated uncertainties.
Additionally, Monte Carlo samples that yield unphysical negative HPE concentrations in any geological reservoir, or HPE concentrations in the LLVPs lower than those in the ambient mantle, are discarded.
Under these conditions, 3683 out of 4000 Monte Carlo samples are accepted (92.1\% acceptance rate).
The resulting median HPE abundances and associated uncertainties for mantle reservoirs are summarized in Table \ref{tab:hpe_abundances}. The full probability distributions of the radiogenic power, HPE abundances, Th/U ratios, and LLVP HPE enrichment factors derived from the accepted samples are shown in Fig. \ref{fig:pdfs}.

\subsection{Parameterization of mantle HPE partitioning}
To characterize the chemical contrast between LLVPs and the ambient mantle and relate the LLVP mass to the bulk mantle mass, we define the LLVP HPE enrichment factor $E^X$ ($X \in \{\mathrm{U, Th}\}$) and the LLVP mass fraction $f_m$, given by
\begin{equation}
    E^X = \frac{A^X_{\mathrm{LLVP}}}{A^X_{\mathrm{amb}}},
\end{equation}
\begin{equation}
    f_m=\frac{m_{\mathrm{LLVP}}}{m_{\mathrm{mantle}}},
\end{equation}
where $A^X_{\mathrm{LLVP}}$ and $A^X_{\mathrm{amb}}$ represent the abundances of element $X$ in LLVPs and the ambient mantle, while $m_{\mathrm{LLVP}}$ and $m_{\mathrm{mantle}}$ denote the total mass of the LLVP domains and the bulk mantle, respectively.
Within the mass-balance framework, the global average mantle abundance of element $X$, $\langle A^X \rangle$, can be expressed as a mass-weighted sum of the two reservoirs according to
\begin{equation} \label{<A>}
\begin{split}
    \langle A^X \rangle &= f_m A_{\mathrm{LLVP}}^X + (1-f_m) A^X_{\mathrm{amb}} \\
    &= (1 + f_m (E^X - 1)) A^X_{\mathrm{amb}}.
\end{split}
\end{equation}

Using these definitions, the ratio of the $X$ budget in LLVPs to the total mantle $X$ budget can be calculated from the same mass-balance relation. 
Specifically,
\begin{equation}
\label{f^X_LLVP}
f^X_\mathrm{LLVP}
= \frac{m^X_{\mathrm{LLVP}}}{m^X_{\mathrm{mantle}}}
= \frac{f_m A^X_{\mathrm{LLVP}}}{\langle A^X \rangle}
= \frac{f_m E^X}{1 + f_m(E^X-1)}.
\end{equation}

By defining LLVP boundaries using the seismic tomographic consensus model, we obtain the corresponding LLVP mass fraction $f_m$ of 3.0\%.
Under our mass-balance framework, this small volume requires placing a large fraction of the mantle HPE budget into the LLVPs, yielding high enrichment factors of $E^\mathrm{U} = 25.2^{+16.3}_{-13.3}$ and $E^\mathrm{Th} = 36.6^{+31.1}_{-19.7}$.
These values imply that the LLVPs contain $43.8^{+12.4}_{-16.9}\%$ and $53.1^{+14.6}_{-18.8}\%$ of the total mantle U and Th budgets, respectively.
The quoted uncertainties represent the central 68\% intervals.
We emphasize that this baseline heterogeneous configuration represents a highly enriched end-member scenario, derived mathematically from combining a small consensus LLVP geometry with mass-balance constraints, rather than a universally representative geochemical model.
Consequently, the positive sensitivity results established for our baseline benchmark illustrate detector capabilities under a strong end-member regime.
While these specific values serve as our baselines, we explore a broader range of HPE enrichment factors in Sect. \ref{sec:sensitivity_fHPE}, where the factors are systematically varied from 1.1 to 100.

\section{Geoneutrino Flux Formalization}
\label{sec:methods}

\subsection{General Geoneutrino Flux Integration}
The expected geoneutrino event rate at any given detector position $\bm{r}$ is computed by integrating the spatial distribution of HPEs in the Earth.
Because only geoneutrinos from the $^{238}\mathrm{U}$ and $^{232}\mathrm{Th}$ decay chains have energies above the 1.8~MeV kinematic threshold of the IBD reaction \eqref{ibd} \citep{raghavan1998measuring}, our calculation considers exclusively these two isotopes.
The expected geoneutrino signal $S(\bm{r})$ is calculated by integrating over both the antineutrino energy $E_\nu$ and the Earth volume $V_\oplus$ according to
\begin{equation} \label{flux_formula0}
\begin{split}
S(\bm{r}) = & \, \varepsilon n_p \sum_{X \in \{\mathrm{U, Th}\}} \frac{N_A \lambda^X}{\mu^X} \int_{E_\nu} \sigma_{\mathrm{IBD}} (E_\nu) \frac{d n^X(E_\nu)}{d E_\nu} d E_\nu \\
& \times \int_{V_\oplus} \frac{\rho(\bm{r}') A^X(\bm{r}') P_{\mathrm{ee}}(E_\nu, |\bm{r} - \bm{r}'|)}{4\pi |\bm{r} - \bm{r}'|^2} d \bm{r}',
\end{split}
\end{equation}
where $\varepsilon$ is the detection efficiency, which is set to unity throughout this study as a benchmark assumption; a different efficiency would simply rescale the expected event rate.
$N_A$ is Avogadro's constant, and $n_p$ represents the total target proton number in the detector, which can be derived by
\begin{equation} \label{proton_num}
    n_p = V \rho Y_p,
\end{equation}
\noindent where $V$ is the fiducial volume of the detector, $\rho$ is mass density of the liquid scintillator, and $Y_p$ is the number of target protons per unit mass of the liquid scintillator.
The variables $\lambda^X$ and $\mu^X$ denote the decay constant and molar mass of isotope $X$, respectively.
The energy integrand combines the differential geoneutrino energy spectrum $d n^X / d E_\nu$ per decay chain of isotope $X$ with the IBD cross section $\sigma_{\mathrm{IBD}}$, which quantifies the probability of an incoming geoneutrino interacting with a target proton in the detector to produce a detectable signal.
The spatial integrand accounts for the local rock density $\rho(\bm{r}')$ and the specific isotope abundance $A^X(\bm{r}')$ at the source location $\bm{r}'$.
Neutrino oscillations, which convert a fraction of electron antineutrinos into other flavors during propagation \citep{fukuda1998evidence}, are incorporated through the survival probability $P_{\mathrm{ee}}$.
All fundamental physical constants adopted for this calculation are explicitly defined and referenced in Table \ref{tab:isotope_constants}.

\begin{table*}[htb]
    \centering
    \caption{Summary of physical constants and reference models used in the geoneutrino flux calculation. Global parameters apply universally to the calculation while isotope-specific values are provided for the distinct decay chains.}
    \label{tab:isotope_constants}
    \setlength{\tabcolsep}{8pt}
    \renewcommand{\arraystretch}{1.2}
    \begin{tabular}{llccc}
        \toprule
        Parameter & Symbol & $^{238}\mathrm{U}$ & $^{232}\mathrm{Th}$ & Reference \\
        \midrule
        \textit{Isotope-specific properties} & & & & \\
        Decay constant & $\lambda^X$ & $1.55\times10^{-10}$ yr$^{-1}$ & $4.92\times10^{-11}$ yr$^{-1}$ & \cite{mcdonough2020radiogenic} \\
        Molar mass & $\mu^X$ & $238.03$ g/mol & $232.04$ g/mol & \cite{mcdonough2020radiogenic} \\
        Heat production rate & $H^X$ & $9.49\times10^{-5}$ W/kg & $2.62\times10^{-5}$ W/kg & \cite{mcdonough2020radiogenic} \\
        Antineutrinos per chain & $\bar{\nu}_e$ & $6$ & $4$ & \cite{enomoto2005phd} \\
        $\bar{\nu}_e$ energy spectrum & $dn^X/dE_\nu$ & \multicolumn{2}{c}{Energy-dependent function} & \cite{enomoto2005phd} \\
        \midrule
        \textit{Global physics parameters} & & \multicolumn{2}{c}{Value or Description} & \\
        Avogadro's constant & $N_A$ & \multicolumn{2}{c}{$6.022\times10^{23}$ mol$^{-1}$} & \cite{mohr2025codata} \\
        IBD cross section & $\sigma_{\mathrm{IBD}}$ & \multicolumn{2}{c}{Energy-dependent function} & \cite{strumia2003ibd} \\
        $\bar{\nu}_e$ survival probability & $P_{\mathrm{ee}}$ & \multicolumn{2}{c}{Quantum mechanical effect} & \cite{PDF2024cfk} \\
        Target proton number & $n_p$ & \multicolumn{2}{c}{Detector-specific scalar} & \\
        \bottomrule
    \end{tabular}
\end{table*}

While Eq. \eqref{flux_formula0} defines the total signal integrated over all angles, directional geoneutrino detection necessitates a differential approach to the source geometry.
When the incoming direction of geoneutrinos is taken into account, instead of integrating over the entire Earth volume $V_\oplus$ as a single scalar quantity, we evaluate the event rate as a function of incoming angle by isolating the radiogenic contributions from specific directions.

\subsection{Linear Superposition of Signals and Inversion Parameterization}
\label{inversionParam}
By separating Eq. \eqref{flux_formula0} into different regions of the Earth, the total observed signal at a detector can be expressed as a linear superposition of crustal contributions $S_{\mathrm{crust}}$, mantle components, and non-geoneutrino backgrounds $S_{\mathrm{BG}}$ (e.g., reactor neutrinos, ($\alpha$, n) reactions and accidental coincidences), written schematically as
\begin{equation}
S_{\mathrm{total}}(\bm{r}) = S_{\mathrm{BG}}(\bm{r}) + S_{\mathrm{crust}}(\bm{r}) + \sum_{\substack{i \in \mathrm{mantle} \\ X \in \{\mathrm{U, Th}\}}} A^X_i \frac{\partial S^X_i(\bm{r})}{\partial A^X_i}.
\end{equation}
Here, the variable $A^X_i$ represents the abundance of the element $X$ in the $i$-th mantle component, while $\partial S^X_i/\partial A^X_i$ represents the corresponding signal contribution per unit abundance of $X$, which serves as the signal template for the fitting.
While previous studies have successfully used this framework to constrain the radiogenic power in the bulk mantle \citep{fiorentini2012mantle,vsramek2016revealing,sammon2022quantifying}, they inherently treated the mantle as a single geochemically homogeneous reservoir.

To test whether geoneutrino detections can extract signals from LLVPs, we extend this linear framework to a mantle model consisting of LLVPs and the ambient mantle.
Depending on the target parameters, the total signal can be formulated in two distinct ways. 
If the primary goal is to directly fit for the HPE abundances of the individual mantle domains, the total signal measured is expressed as a linear combination of the background, crustal, and specific mantle-domain contributions according to
\begin{equation} \label{AAori}
\begin{split}
S_{\mathrm{total}}(\bm{r}) = & \, S_{\mathrm{BG}}(\bm{r}) + S_{\mathrm{crust}}(\bm{r}) \\
& + \sum_{X\in\{\mathrm{U, Th}\}} \left( A^X_\mathrm{LLVP} \frac{\partial S^X_\mathrm{LLVP}(\bm{r})}{\partial A^X_\mathrm{LLVP}} + A^X_\mathrm{amb} \frac{\partial S^X_\mathrm{amb}(\bm{r})}{\partial A^X_\mathrm{amb}} \right).
\end{split}
\end{equation}

In experimental practice, a widely adopted approach in geoneutrino analysis is to fix the $\mathrm{Th/U}$ mass ratio during spectral fitting (e.g., \citet{borexino2020geonu,kamland2022geonu,juno2025geonu}).
This approach accounts for the significant overlap between the energy spectra and the limited energy resolution of the detector.
Under the assumption of fixed $\mathrm{Th/U}$ ratios in LLVPs and the ambient mantle, Eq.~\eqref{AAori} simplifies to

\begin{equation} \label{AA}
S_{\mathrm{total}}(\bm{r}) = S_{\mathrm{BG}}(\bm{r}) + S_{\mathrm{crust}}(\bm{r})
+ A^\mathrm{U}_\mathrm{LLVP} c_\mathrm{LLVP}(\bm{r}) + A^\mathrm{U}_\mathrm{amb} c_\mathrm{amb}(\bm{r}),
\end{equation}

\noindent where $c_\mathrm{LLVP}(\bm{r})$ and $c_\mathrm{amb}(\bm{r})$ denote the effective response coefficients per unit U abundance for the LLVPs and ambient mantle, respectively.
They are defined as
\begin{equation} \label{eq:c_response_def}
c_j(\bm{r}) = \frac{\partial S^\mathrm{U}_j(\bm{r})}{\partial A^\mathrm{U}_j} + R_j \frac{\partial S^\mathrm{Th}_j(\bm{r})}{\partial A^\mathrm{Th}_j}, \quad \quad (j \in \{\mathrm{LLVP}, \mathrm{amb}\})
\end{equation}
\noindent with $R_j = A^\mathrm{Th}_j / A^\mathrm{U}_j$ representing the prescribed $\mathrm{Th/U}$ mass ratio in reservoir $j$.
By doing so, the number of free parameters in the fit is effectively reduced from four ($A^\mathrm{U}_\mathrm{LLVP}$, $A^\mathrm{Th}_\mathrm{LLVP}$, $A^\mathrm{U}_\mathrm{amb}$, $A^\mathrm{Th}_\mathrm{amb}$) to two ($A^\mathrm{U}_\mathrm{LLVP}$, $A^\mathrm{U}_\mathrm{amb}$).

When geoneutrino measurements provide only event rates and lack directional information, a single detector provides only one observable, rendering the two-parameter system ($A^\mathrm{U}_\mathrm{LLVP}$ and $A^\mathrm{U}_\mathrm{amb}$) mathematically underconstrained. 
Resolving these parameters requires either directional sensitivity at a single detector or a distributed network of rate-only detectors. 
In either case, breaking parameter degeneracy fundamentally relies on sufficient contrast in the signal response ratio $c_\mathrm{LLVP}/c_\mathrm{amb}$ across the set of observables.
If the two reservoirs have similar signal responses, changes in their abundances produce nearly indistinguishable changes in the observables, preventing their individual contributions from being separated.
This contrast can arise either spatially across detector sites or directionally across arrival vectors, providing useful information to distinguish the two reservoirs.

Alternatively, we can analytically recast Eq. \eqref{AA} to evaluate whether the measurement can jointly constrain the global mean mantle U abundance $\langle A^\mathrm{U} \rangle$ and the LLVP U enrichment factor $E^\mathrm{U}$.
By substituting Eq. \eqref{<A>} into Eq. \eqref{AA}, the total measured signal can be re-parameterized as a function of $\langle A^\mathrm{U} \rangle$ and $E^\mathrm{U}$:
\begin{equation} \label{Af}
\begin{split}
S_{\mathrm{total}}(\bm{r}) = & \, S_{\mathrm{BG}}(\bm{r}) + S_{\mathrm{crust}}(\bm{r}) \\
& + \frac{\langle A^\mathrm{U} \rangle}{1 + f_m (E^\mathrm{U}-1)} \left( c_\mathrm{amb}(\bm{r}) + E^\mathrm{U} c_\mathrm{LLVP}(\bm{r}) \right).
\end{split}
\end{equation}

\section{Geoneutrino Detection with Angular Resolution}
\label{sec:detector}

\subsection{Experimental Framework}
\label{expframe}
To effectively detect geoneutrinos originating in the mantle, it is crucial to minimize the overwhelming neutrino background emitted by the continental crust and nuclear reactors.
Oceanic environments present an optimal geological setting for mantle geoneutrino measurements, because the reduced thickness \citep{mooney2023ecm} and intrinsically low HPE concentrations \citep{white2014} of the oceanic crust substantially suppress local emissions.
This condition allows the mantle contribution to exceed 70\% of the total expected geoneutrino flux \citep{sramek2013geophysical,huang2013reference}.

To quantitatively evaluate the potential of directional geoneutrino detection in such an environment, we use the proposed Ocean Bottom Detector (OBD) project \citep{obd2026whitepaper}, a 1.5- or 17-kt liquid scintillator neutrino detector planned for deployment in the Pacific Ocean, as a representative experimental framework.
For our sensitivity study, we simulate a 1.5-kt OBD situated at 15$^\circ$S, 150$^\circ$W in the South Pacific.
This specific location sits directly above the Pacific LLVP and maximizes the distance from continental margins (Fig. \ref{fig:U_distribution}), serving as an ideal testing ground for our proposed observational strategy. 

To ensure a more rigorous sensitivity study and avoid overly optimistic results, we adopt the detector parameters and non-geoneutrino background levels reported in \cite{sakai2021obd} and \cite{watanabe2023obd}.
These prior studies assumed that the internal radioactive impurities of the OBD are equivalent to those of the KamLAND during its initial experimental phase \citep{kamland2015solarnu}, resulting in a relatively high background level.
Based on the signal-to-noise ratio optimization, the fiducial volume of the 1.5-kt OBD can be approximated by a cylindrical volume with a diameter of 10.6 m and a height of 12.6 m.
We adopt a density of 863 kg/m$^3$ for the liquid scintillator based on linear alkylbenzene (LAB), accounting for the effects of temperature and pressure, and a proton number of $7.26\times10^{28}$ per kg of scintillator.
According to Eq. \eqref{proton_num}, these parameters correspond to $7.0\times10^{31}$ protons within the fiducial volume of the detector.

Furthermore, to model rigorously the external reactor antineutrino background, we use actual 2024 operational data from the International Atomic Energy Agency (IAEA) integrated into the global modeling framework of \cite{dye2015global}.
The corresponding reactor isotope emission spectra are constructed according to the refined flux models presented by \cite{giunti2022reactor}.
Adopting these specific internal background assumptions alongside the most recent empirical reactor data ensures a robust and conservative evaluation of our directional detection sensitivity.

\subsection{Angular Resolution in Geoneutrino Detectors}
\label{angres}
Geoneutrinos are detected via the IBD reaction \eqref{ibd}, which yields a prompt positron ($e^+$) annihilation signal and a delayed neutron ($n$) capture event.
Due to reaction kinematics, the neutron is preferentially emitted in the forward direction, meaning the spatial vector connecting the prompt and delayed interaction vertices theoretically traces the incident antineutrino's trajectory \citep{vogel1999angular}.
However, in standard liquid scintillators used by decommissioned and existing geoneutrino detectors, the prolonged random walk of the neutron smears out this initial displacement, resulting in a loss of directional information \citep{tanaka20146li}.

To overcome this fundamental limitation, recent detector advancements introduce three critical enhancements.
Chemically, doping the scintillator medium with isotopes that possess exceptionally high neutron capture cross sections, such as $^{6}\mathrm{Li}$ or Gd, drastically reduces the number of neutron scattering events \citep{tanaka20146li,seo2020labls}.
Optically, water-based or slow liquid scintillators enable directional determination by decoupling directional Cherenkov photons from isotropic scintillation light \citep{wang2020hunting,sun2026potassium}.
Structurally, dividing the continuous detector volume into segmented lattices substantially improves the spatial vertex resolution \citep{safdi2015directional,sutanto2021sandd}.
Complementing these physical enhancements, statistical algorithms have recently been developed to optimally extract the underlying angular distribution from the reconstructed vertices \citep{yepez2026algorithm}.
This powerful synergy of hardware and software makes it possible to constrain the incoming angle $\theta$ of geoneutrinos with an effective angular resolution $\sigma_\theta$ in the future \citep{duvall2024directional,crow2026enhancing}.

To assess the detector performance required to extract LLVP signals from the ambient mantle, we calculate the signal event rate as a function of the incoming arrival angle $\theta$ using Eq. \eqref{flux_formula0}.
The angular resolution $\sigma_\theta$ adopted in this study is not intended to represent demonstrated performance of an existing OBD design.
We treat $\sigma_\theta$ as a target detector-performance parameter and evaluate how the sensitivity to LLVP signals varies with this parameter.
For each target value of $\sigma_\theta$, we model the corresponding directional response by applying a Gaussian smearing over the spherical domain defined by the nadir and azimuthal angles.
The smearing is defined in terms of the angular separation between the true and reconstructed directions, with $\sigma_\theta$ representing the standard deviation of this separation.
Thus, the assumed values of $\sigma_\theta$ should be interpreted as target performance levels for a future directional geoneutrino detector rather than as demonstrated detector performance.

\begin{figure*}[!htbp]
    \centering
    \includegraphics[width=0.75\textwidth]{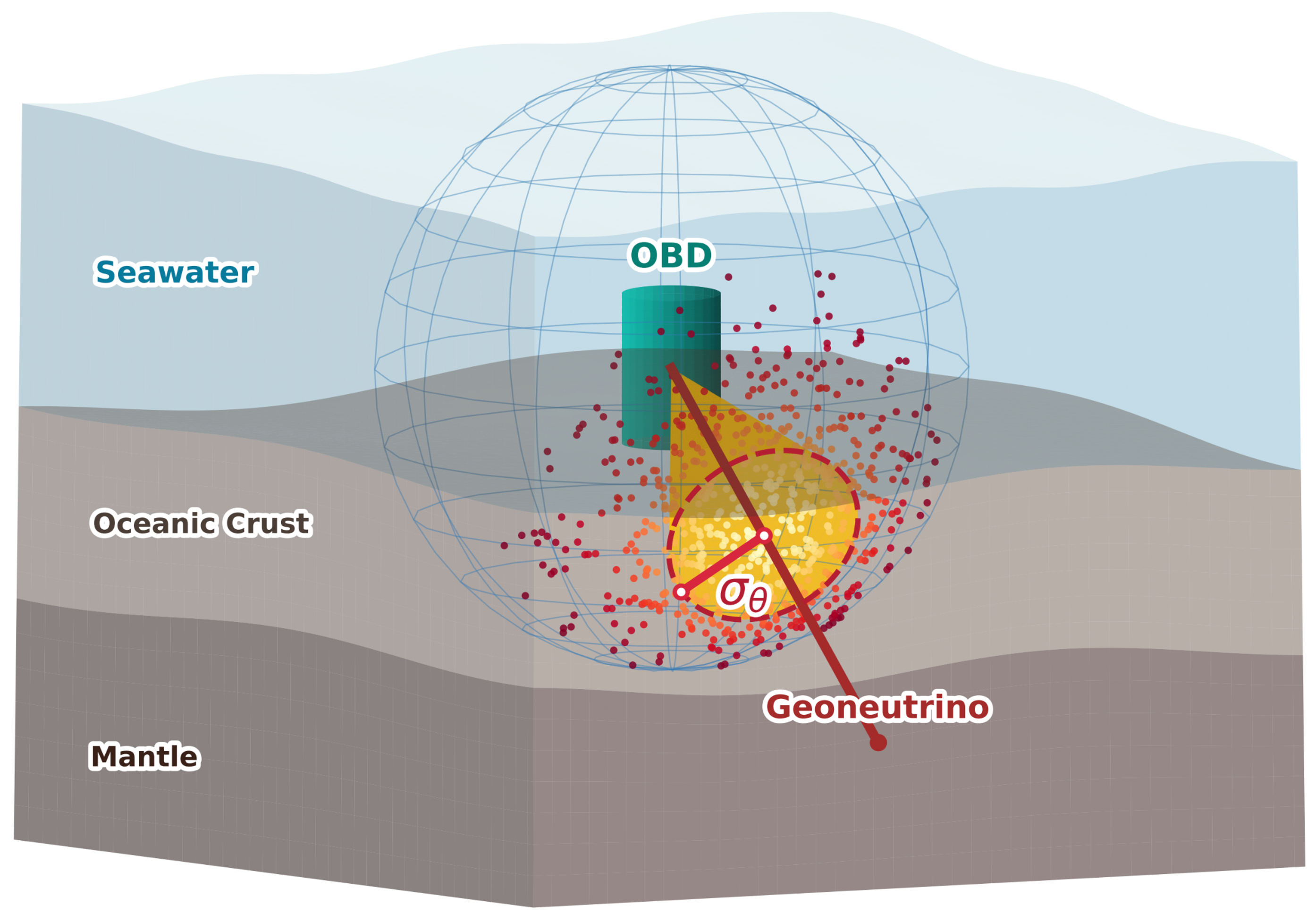}
    \caption{Schematic depiction of directional geoneutrino detection using an Ocean Bottom Detector (OBD) (not to scale). 
    If geoneutrinos arrive at the OBD from the same direction, their reconstructed arrival directions are distributed around the input direction due to the detector's finite angular resolution ($\sigma_\theta$), as illustrated by the scattered points on the unit sphere.
    The angular separation between the input and reconstructed directions follows a spherical Gaussian distribution, with $\sigma_\theta$ denoting the standard deviation of this angular distribution.
    The illustrated angular resolution represents a target performance benchmark for a future directional geoneutrino detector and does not imply demonstrated performance of an existing detector design.}
    \label{fig:detector}
\end{figure*}

Analogous to the linear parameterizations defined in Eqs. \eqref{AA} and \eqref{Af}, the angular distribution of the total observed event $S_\mathrm{total}(\bm{r}, \theta)$ is evaluated simply as a linear superposition of the directionally smeared contributions from the crust ($S_\mathrm{crust}(\bm{r}, \theta)$), the ambient mantle ($S_\mathrm{amb}(\bm{r}, \theta)$), the LLVPs ($S_\mathrm{LLVP}(\bm{r}, \theta)$), and the non-geoneutrino backgrounds ($S_\mathrm{BG}(\bm{r}, \theta)$).
This direct linear combination principle applies equally when transforming the basis variables to the global average mantle U abundance $\langle A^\mathrm{U} \rangle$ and the LLVP U enrichment factor $E^\mathrm{U}$.

\section{Testing for HPE Enrichment in LLVPs via Directional Geoneutrino Detection}
\label{sec:directional_solution}
In this chapter, we present the expected sensitivity of directional geoneutrino detection to HPE enrichment in LLVPs in a stepwise manner.
In Sect. \ref{sec:angulardist}, we first calculate the event rates and nadir-angle distributions of signals, illustrating how the localized LLVP contribution produces directional information.
Then, in Sect. \ref{directional_OBD}, we use the median values of the Earth model parameters as a representative case to quantify how a direction-sensitive OBD can constrain the HPE abundances of LLVPs and the ambient mantle.
Next, in Sect. \ref{network}, we compare these constraints with those obtained from a global network of past, existing, under-construction, and planned rate-only detectors, both with and without an additional rate-only OBD, thereby assessing the additional information provided by directional sensitivity.
Finally, in Sect. \ref{sec:sensitivity_fHPE}, to provide a more general assessment, we take into account uncertainties in the Earth model, backgrounds, and statistical fluctuations in the event counts using a Monte Carlo method, and evaluate the statistical significance for rejecting a homogeneous mantle hypothesis as a function of angular resolution, experimental duration, and HPE enrichment factors.

\subsection{Expected Event Rates and Angular Distributions of Signals}
\label{sec:angulardist}
\begin{figure*}[!htbp]
    \centering
    \includegraphics[width=0.85\textwidth]{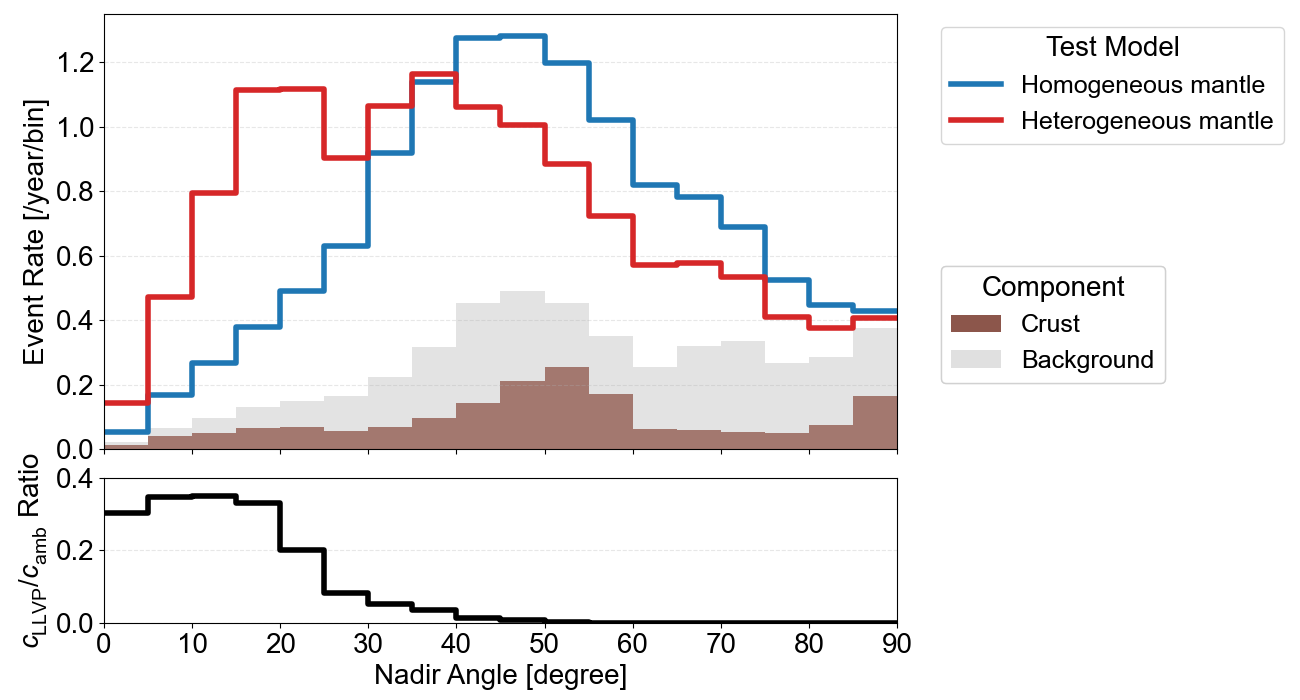}
    \caption{Predicted nadir-angle distributions of geoneutrino and background event rates for a 1.5-kt Ocean Bottom Detector (OBD) in the South Pacific, binned in $5^\circ$ intervals. 
    The solid lines represent the total event rates for the chemically homogeneous mantle model (blue) and the heterogeneous mantle model, where the LLVP geometry is defined by the tomographic consensus model with $E^\mathrm{U}=25.2$ and $E^\mathrm{Th}=36.6$ (red). 
    The filled areas indicate the contributions from the crust (brown) and background sources (light gray), which are common to both models. 
    The bottom panel shows the nadir-angle distribution of the $c_{\mathrm{LLVP}}/c_{\mathrm{amb}}$ ratio for the heterogeneous test model.}
    \label{fig:nadir_distribution}
\end{figure*}

As our primary aim is to provide an initial assessment of whether directional sensitivity can probe the modeled LLVP signals, we integrate over the azimuthal angle rather than resolving it, avoiding the need for additional binning optimization while retaining sufficient statistics in each bin.
Fig. \ref{fig:nadir_distribution} shows the nadir-angle distribution of the signals before incorporating detector's angular resolution.
In this convention, $0^\circ$ represents neutrinos coming from directly below (the Earth's center), while $90^\circ$ represents those arriving from the horizon.
Here, the non-geoneutrino background signals are assumed to be isotropically distributed over the unit sphere, as these backgrounds are not expected to exhibit a preferred arrival direction.

Even when accounting for the comprehensive background contributions, the localized HPE enrichment in LLVPs manifests prominently in the heterogeneous models, particularly at low nadir angles directly beneath the detector.
This spatial distinction occurs because the anomalous signal originating from the HPE-enriched LLVPs is highly concentrated within the low nadir-angle region directly beneath the detector.
However, as summarized in Table \ref{tab:event_rates}, the predicted total event rates at the South Pacific site vary by about 4\% between the homogeneous and heterogeneous models, corresponding to a difference of only $\sim$0.6 event/yr.
This indistinguishability once again demonstrates that rate-only measurements inherently lack the sensitivity to resolve deep-mantle chemical heterogeneities.
Consequently, directional geoneutrino detection becomes necessary to distinguish a homogeneous mantle from one containing enriched thermochemical piles.

\begin{table*}[htb]
    \centering
    \caption{Predicted geoneutrino and non-geoneutrino background event rates (events/yr) for a 1.5-kt OBD under each test model. Non-geoneutrino background including reactor neutrinos, ($\alpha$, n) reactions, and accidental signals from detector impurities.}
    \label{tab:event_rates}
    \setlength{\tabcolsep}{4pt}
    \renewcommand{\arraystretch}{1.1}
    \begin{tabular}{lccccccc}
        \toprule
        Test model & Crust & Ambient mantle & LLVP domains & Reactor & ($\alpha$, n) reaction & Accidental & Total \\
        \midrule
        Homogeneous & $1.9 \pm 0.4$ & $8.0 \pm 1.8$ & -- & $0.6 \pm 0.1$ & $3.0 \pm 0.5$ & $1.9 \pm 0.1$ & $16.1 \pm 1.8$ \\
        Heterogeneous & $1.9 \pm 0.4$ & $4.1 \pm 0.5$ & $4.5 \pm 2.2$ & $0.6 \pm 0.1$ & $3.0 \pm 0.5$ & $1.9 \pm 0.1$ & $16.7 \pm 2.2$ \\
        \bottomrule
    \end{tabular}
\end{table*}

\subsection{Constraining HPE Enrichment in LLVPs with Angular Profiles}
\label{directional_OBD}
Building upon the formulation described in Sect. \ref{inversionParam}, we evaluate how precisely the HPE abundances of LLVPs and the ambient mantle can be constrained using directional geoneutrino measurements.
To characterize the expected precision under a representative realization of the Earth model, following the statistical framework described in Appendix \ref{app:statistics}, we use the median HPE abundances of the heterogeneous test model rather than propagating their uncertainties in this section.
We construct a binned Poisson likelihood for the observed event counts in each angular bin.
The likelihood is evaluated under both heterogeneous and homogeneous mantle assumptions, while uncertainties in backgrounds are incorporated through Gaussian penalty terms.

As an illustrative example, we consider an optimistic scenario with an advanced angular resolution of $\sigma_\theta = 20^\circ$ to explore the potential of directional information.
We assume that the OBD operates for 20 years, consistent with the operating period of KamLAND \citep{kamland2022geonu}, and fix the Th/U ratios of both the LLVPs and the ambient mantle to their respective median values listed in Table \ref{tab:hpe_abundances} during the fitting.
Using the angular templates for the homogeneous and heterogeneous mantle models in Eqs. \eqref{AA} and \eqref{Af}, we fit the HPE concentrations under the two corresponding model assumptions.

The resulting constraints on the mantle U abundances and LLVP U enrichment factor are summarized in Table \ref{tab:fit_result}.
Fitting the data with the homogeneous mantle template yields $\langle A^\mathrm{U} \rangle = 16.8^{+1.7}_{-1.6}$ ppb, substantially higher than the value prescribed in the test model, 13.9 ppb.
This reflects the inability of the homogeneous model to account for the localized HPE enrichment within the LLVPs, which is instead absorbed into the inferred bulk mantle abundance.
In contrast, the heterogeneous template independently recovers U abundances in both the ambient mantle and LLVPs, yielding constraints consistent with their respective values.

\begin{table*}[htb]
    \centering
    \caption{Expected constraints on mantle U abundances and LLVP enrichment factor for different geoneutrino measurement configurations.}
    \label{tab:fit_result}
    \begin{tabular}{lcccc}
        \toprule
        & $A^\mathrm{U}_\mathrm{amb}$ (ppb) 
        & $A^\mathrm{U}_\mathrm{LLVP}$ (ppb) 
        & $\langle A^\mathrm{U} \rangle$ (ppb) 
        & $E^\mathrm{U}$ \\
        \midrule
        Prescribed value 
        & $8.0$ & $202.7$ & $13.9$ & $25.2$ \\
        \midrule
        \multicolumn{5}{l}{\textit{Single directional OBD (with $\sigma_\theta=20^\circ$ for $T=20$ yr)}} \\
        \quad Fit with homogeneous template
        & -- & -- & $16.8^{+1.7}_{-1.6}$ & -- \\
        \quad Fit with heterogeneous template
        & $8.0^{+2.5}_{-2.4}$
        & $202.7^{+48.3}_{-46.8}$
        & $13.9 \pm 1.7$
        & $25.2^{+17.9}_{-9.9}$ \\
        \midrule
        \multicolumn{5}{l}{\textit{Rate-only network with 10 detectors (under idealized conditions)}} \\
        \quad Fit with homogeneous template
        & -- & -- & $11.8 \pm 3.5$ & -- \\
        \quad Fit with heterogeneous template
        & $8.0 \pm 7.4$
        & $202.7$ (unconstrained)
        & $13.9^{+8.2}_{-5.6}$
        & $25.2$ (unconstrained) \\
        \midrule
        \multicolumn{5}{l}{\textit{Rate-only network with 10 detectors and a rate-only OBD (under idealized conditions)}} \\
        \quad Fit with homogeneous template
        & -- & -- & $15.2 \pm 1.8$ & -- \\
        \quad Fit with heterogeneous template
        & $8.0^{+6.5}_{-6.6}$
        & $202.7^{+165.3}_{-164.0}$
        & $13.9 \pm 2.1$
        & $25.2^{+222.3}_{-23.3}$ \\
        \bottomrule
    \end{tabular}
\end{table*}

\begin{figure*}[!htbp]
    \centering
    \includegraphics[width=\textwidth]{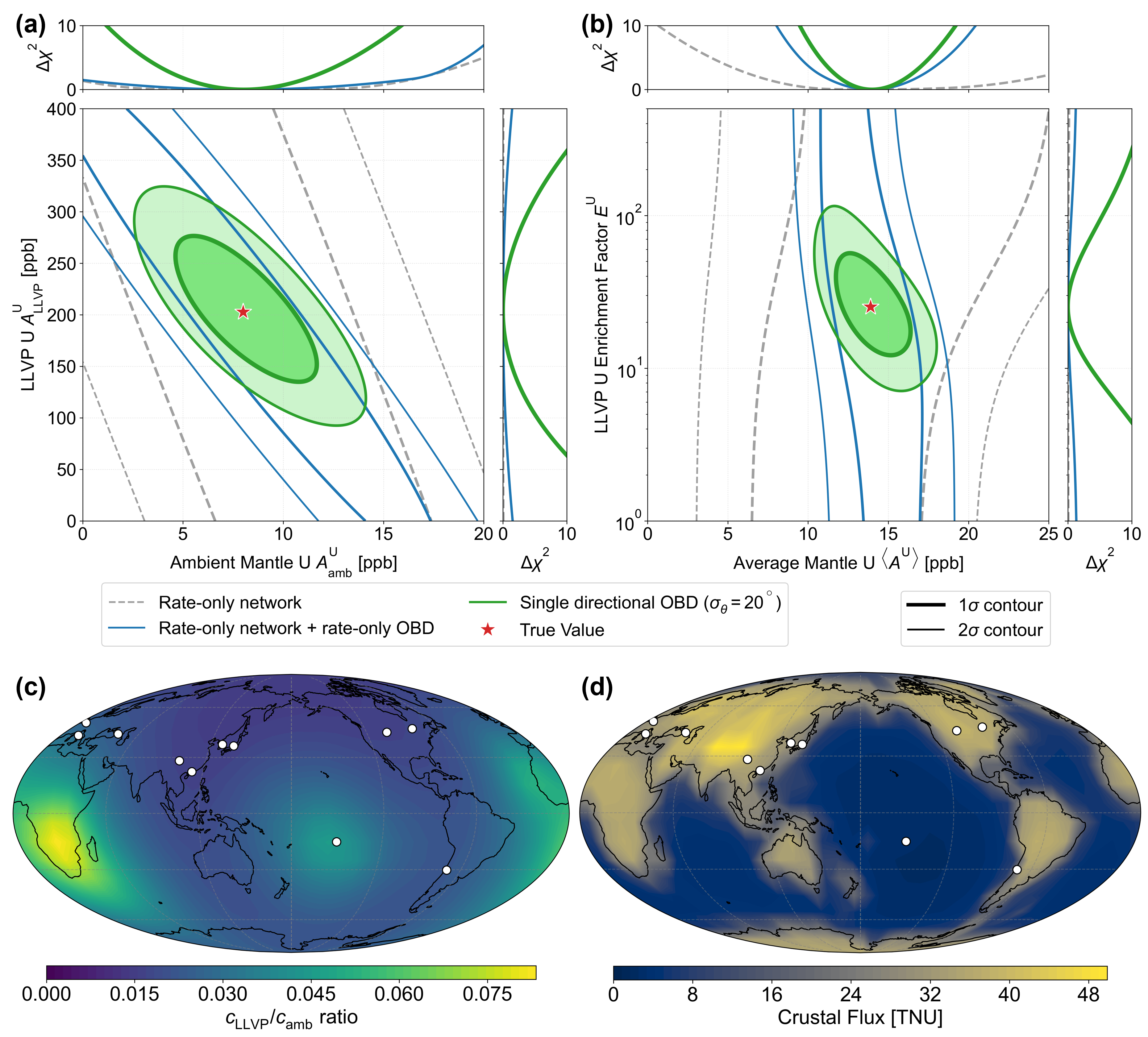}
    \caption{Constraints on mantle U abundances from rate-only and directional geoneutrino detection.
    In (a) and (b), green contours show the constraints from a single directional Ocean Bottom Detector (OBD) in the South Pacific with an angular resolution of $\sigma_\theta = 20^\circ$ over a 20-year experimental period.
    Gray dashed contours show the constraints from the global network of 10 rate-only geoneutrino detectors under the idealized conditions described in Sect. \ref{network}, while blue contours show the constraints obtained by supplementing this network with a rate-only OBD.
    For each configuration, the thin and thick contours represent the $1\sigma$ and $2\sigma$ confidence levels, respectively.
    (a) Confidence regions in the ambient mantle U abundance ($A_\mathrm{amb}^\mathrm{U}$) versus LLVP U abundance ($A_\mathrm{LLVP}^\mathrm{U}$) plane.
    (b) Equivalent confidence regions in the bulk average mantle U abundance ($\langle A^\mathrm{U} \rangle$) versus LLVP U enrichment factor ($E^\mathrm{U}$) plane.
    (c) Global spatial distribution of the $c_\mathrm{LLVP}/c_\mathrm{amb}$ ratio.
    (d) Global spatial distribution of the crustal geoneutrino flux, expressed in Terrestrial Neutrino Unit (TNU), where 1 TNU corresponds to 1 event per $10^{32}$ target protons per year.
    The white circles in (c) and (d) indicate the locations of the rate-only geoneutrino detectors used to construct the measurement configurations shown in (a) and (b).}
    \label{fig:OBD_chi2}
\end{figure*}

The resulting constraints are further illustrated by the green confidence contours in the $(A^\mathrm{U}_\mathrm{amb}, A^\mathrm{U}_\mathrm{LLVP})$ parameter space (Fig. \ref{fig:OBD_chi2}a), the confidence contours are fully closed within the evaluated parameter range, demonstrating that directional geoneutrino measurements can independently constrain the HPE abundances of LLVPs and the ambient mantle.
The constraints can be equivalently expressed in terms of the global average mantle U abundance $\langle A^\mathrm{U} \rangle$ and the LLVP U enrichment factor $E^\mathrm{U}$ (Fig. \ref{fig:OBD_chi2}b), providing a direct characterization of both the overall mantle U abundance and its spatial heterogeneity.

These results indicate that, within the case considered here, directional geoneutrino measurements can provide sensitivity to HPE enrichment in LLVPs relative to the ambient mantle.
Future studies extending this framework to alternative models, such as HPE-enriched D$''$-layer model, will be important for assessing the capability of directional geoneutrino measurements to distinguish between different deep-mantle structures.

\subsection{Comparison with Rate-only Detection Network}
\label{network}
In order to assess how much directional information improves the ability to distinguish the HPE abundances of LLVPs and the ambient mantle, we compare the constraints from directional geoneutrino detection with those achievable using a global network of rate-only detectors.
We consider a global network comprising 10 geoneutrino detectors worldwide, including those from past, present, and under-construction experiments: KamLAND \citep{kamland2022geonu}, Borexino \citep{borexino2020geonu}, SNO+ \citep{sno2026geonu}, JUNO \citep{juno2025geonu}, and JNE \citep{vsramek2016revealing}, as well as planned experiments: THEIA \citep{theia2022geonu}, SuperChooz \citep{liquido2026geonu}, ANDES \citep{ANDES2012potential}, Baksan \citep{baksan2017whitepaper}, and $\nu\mathrm{EYE}$ \citep{nueye2026whitepaper}.

Because these detectors differ substantially in size, geographic location, overburden, and internal radioactive contamination, a detailed detector-by-detector treatment would introduce additional assumptions that are not directly relevant to this comparison.
We therefore adopt a simplified and deliberately optimistic treatment for the rate-only network.
Specifically, we assume a completely background-free environment ($S_\mathrm{BG}(\bm{r})=0$), together with an optimistic 10\% uncertainty on the measured total geoneutrino flux $S_\mathrm{total}(\bm{r})$ and a stringent 10\% uncertainty on the crustal geoneutrino flux $S_\mathrm{crust}(\bm{r})$ for every detector in this rate-only network.
These assumptions are substantially more optimistic than current state-of-the-art experimental ($\sim 16 \%$, \citet{kamland2022geonu}) and crustal geoneutrino flux modeling uncertainties ($\sim 20 \%$, \citet{wipperfurth2020reference}).
The resulting constraints for the rate-only network should therefore be regarded as optimistic estimates of its achievable sensitivity, providing a favorable benchmark against which to assess the additional information gained from directional detection discussed in Sect. \ref{directional_OBD}.

Using the same Earth models and analysis method as described in Sect. \ref{directional_OBD}, we evaluate the constraints can be obtained from the rate-only network of 10 detectors.
The resulting constraints on the mantle U abundances and HPE enrichment factor are summarized in Table \ref{tab:fit_result}.
Compared with the single direction-sensitive OBD, the rate-only network yields substantially broader uncertainties.
In particular, the LLVP U abundance $A^\mathrm{U}_\mathrm{LLVP}$ and the LLVP U enrichment factor $E^\mathrm{U}$ cannot be constrained at the $1\sigma$ level.
Note that although the median values of the fitted parameters are consistent with the prescribed values, this agreement is expected because the fitting templates are constructed using the median values of the same Earth model parameters adopted to generate the test model in this section.
Given the large uncertainties of the rate-only constraints, this agreement should therefore not be interpreted as evidence for precise recovery of the LLVP compositions.

The corresponding likelihood contours are shown as gray dashed curves in Fig. \ref{fig:OBD_chi2}.
In both the $(A^\mathrm{U}_\mathrm{amb}, A^\mathrm{U}_\mathrm{LLVP})$ parameter space (Fig. \ref{fig:OBD_chi2}a) and the $(\langle A^\mathrm{U} \rangle, E^\mathrm{U})$ parameter space (Fig. \ref{fig:OBD_chi2}b), the contours form broad, elongated valleys, reflecting a strong parameter degeneracy between the ambient-mantle U abundance and the HPE enrichment of the LLVPs.
This degeneracy prevents the rate-only network from independently constraining the U abundance of LLVPs or their enrichment relative to the ambient mantle, as different combinations of LLVP and ambient-mantle HPE abundances can produce similar total event rates, making it difficult for the rate-only network to identify the contribution from the localized LLVPs.

The origin of this limited constraining power can be understood from the response coefficient ratios and crustal backgrounds of the rate-only detectors.
The response coefficient ratios, $c_\mathrm{LLVP}(\bm{r})/c_\mathrm{amb}(\bm{r})$, for the 10 detectors are shown in Fig. \ref{fig:OBD_chi2}c.
Although the different geographic locations of the detectors introduce some variation in this ratio, its overall variation is only about 0.02.
In contrast, for the direction-sensitive OBD considered in this study, binning the events by nadir angle can produce a larger variation in the response coefficient ratio, reaching up to approximately 0.35 for perfect angular resolution, as illustrated in Fig. \ref{fig:nadir_distribution}.
The angular dependence therefore provides greater leverage for separating the contributions from LLVPs and the ambient mantle than the differences among rate-only detectors.
Fig. \ref{fig:OBD_chi2}d further shows the crustal geoneutrino fluxes at the detector locations.
The OBD has a lower crustal background than the continental detectors, making the deep-mantle contribution more prominent in its total geoneutrino signal.
This lower crustal background further enhances the sensitivity of the OBD to deep-mantle structures.

We further consider a rate-only OBD added to the 10-detector global network described above.
Using the same assumptions of a background-free environment, 10\% uncertainty on the total measured signals, and 10\% uncertainty on the crustal flux, we obtain the fit results summarized in Table \ref{tab:fit_result}.
The additional rate-only OBD improves the constraints on the LLVP parameters, allowing finite $1\sigma$ constraints to be obtained for both $A^\mathrm{U}_\mathrm{LLVP}$ and $E^\mathrm{U}$.
Nevertheless, the resulting uncertainties remain larger than those obtained with a single direction-sensitive OBD.
The corresponding confidence contours are shown in blue in Fig. \ref{fig:OBD_chi2}.
Although the additional rate-only OBD narrows the contours, they remain broadly elongated along the $A^\mathrm{U}_\mathrm{LLVP}$ and $E^\mathrm{U}$ directions.
Thus, even when supplemented by a rate-only OBD, the rate-only detector network considered here, consisting of existing, under-construction, and planned experiments, provides limited constraints on the HPE abundance and enrichment of the LLVPs.

Overall, these results indicate that, a direction-sensitive OBD provides stronger constraints on the HPE contrast between the LLVPs and ambient mantle than rate-only measurements from the detector network considered here, even when supplemented by a rate-only OBD.
We emphasize, however, that this comparison does not constitute a rigorous demonstration that rate-only measurements intrinsically cannot constrain the compositional heterogeneity.
Our rate-only benchmark is based on the network of past, existing, under-construction, and currently planned experiments worldwide described above, together with simplified assumptions adopted for consistency and practicality.
A rigorous assessment of the ultimate capability of rate-only detection would require a systematic optimization of detector locations, overburden, detector masses, experimental periods, and experimental uncertainties, which is beyond the scope of this study.

\subsection{Sensitivity for Excluding a Homogeneous Mantle Model}
\label{sec:sensitivity_fHPE}
To evaluate more generally the capability of directional geoneutrino detection to reject a chemically homogeneous mantle, we quantify the statistical significance at which the homogeneous mantle hypothesis can be excluded under the assumption that the Earth's interior follows the prescribed heterogeneous model.
In this framework, the homogeneous scenario corresponds to HPE enrichment factors of $E^\mathrm{U}=1$ and $E^\mathrm{Th}=1$.

Unlike the parameter constraint analysis in Sect. \ref{directional_OBD}, where the expected angular distributions are constructed using only the median values of the Earth model parameters, here we propagate the uncertainties in the HPE abundances of the Earth models, the crustal geoneutrino flux, and the background components into the significance estimation using a Monte Carlo approach.
For each pseudo-experiment, the number of observed events in each angular bin is sampled from the corresponding Poisson distribution, and the statistical significance is evaluated from a large ensemble of such pseudo-experiments.
Because of the low event statistics and the proximity of the fitted parameters to their physical boundaries, Wilks' theorem \citep{wilks1938} is not assumed to be valid for converting the $\Delta\chi^2$ values into statistical significances.
Instead, we evaluate the significance using the Monte Carlo procedure described in Appendix \ref{app:beyond_Wilks}.
As in the previous analysis, the Th/U ratios are fixed during the fitting process.

\begin{figure*}[!htbp]
    \centering
    \includegraphics[width=0.76\textwidth]{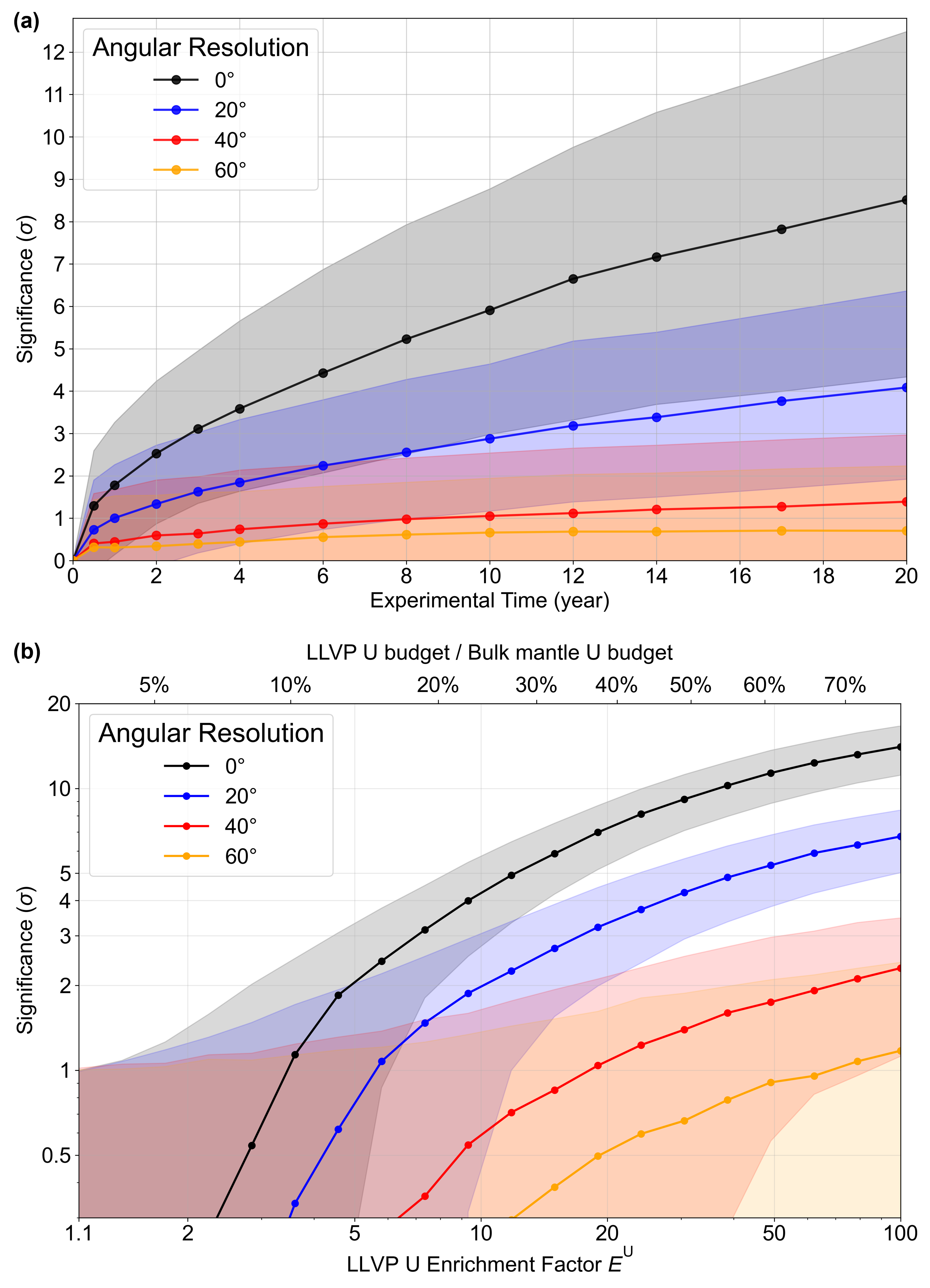}
    \caption{Statistical significance for rejecting a chemically homogeneous mantle hypothesis using a proposed 1.5-kt Ocean Bottom Detector (OBD) in the South Pacific.
    (a) Statistical significance as a function of experimental time for different effective angular resolutions. The heterogeneous mantle test model is characterized by LLVP enrichment factors of $E^\mathrm{U} = 25.2^{+16.3}_{-13.3}$ and $E^\mathrm{Th} = 36.6^{+31.1}_{-19.7}$.
    (b) Statistical significance as a function of the LLVP U enrichment factor $E^\mathrm{U}$ for different detector angular resolutions.
    All calculations in (b) assume a 20-year operational period, and the angular resolutions considered represent benchmark performance levels for a future directional geoneutrino detector, rather than demonstrated detector capabilities.}
    \label{fig:sensitivity_curves}
\end{figure*}

As illustrated in Fig. \ref{fig:sensitivity_curves}a, under the prescribed HPE enrichment factors and LLVP geometry, the exclusion sensitivity depends on both the detector angular resolution $\sigma_\theta$ and the experimental duration $T$.
The solid curves show the median significance obtained from the Monte Carlo ensemble, while the error bands indicate the corresponding central 68\% intervals.
The spread in significance reflects the combined effects of uncertainties in the HPE abundances, the crustal geoneutrino flux, and non-geoneutrino backgrounds, as well as statistical fluctuations in the observed event counts, which are modeled using Poisson sampling.
The sensitivity improves with both increasing experimental duration and improving angular resolution.
At an angular resolution of $\sigma_\theta=20^\circ$, for example, the statistical significance increases from $1.8^{+1.5}_{-1.4}\sigma$ to $4.1^{+2.3}_{-2.2}\sigma$ as the experimental period increases from 4 to 20 years.
From another perspective, for an experimental duration of 20 years, improving the angular resolution from $\sigma_\theta=60^\circ$ to $20^\circ$ increases the statistical significance from $0.7^{+1.5}_{-0.7}\sigma$ to $4.1^{+2.3}_{-2.2}\sigma$.

We further assess the ability of directional geoneutrino detection to reject a chemically homogeneous mantle hypothesis for different assumed values of the LLVP U enrichment factor $E^\mathrm{U}$, given the limited constraints on the chemical composition of the LLVPs from existing geophysical and geochemical observations.
Since $E^\mathrm{U}$ is treated as a free parameter in this analysis, the composition of the ambient mantle cannot be fixed to that of MORB-source, and the MORB-source-based Th/U constraint used in the previous analysis cannot be applied to determine $E^\mathrm{Th}$.
We therefore make the simplifying assumption that U and Th have the same enrichment factor in the LLVPs, i.e., $E^\mathrm{Th}=E^\mathrm{U}$, such that $E^\mathrm{Th}$ is not treated as an independent parameter.

As shown in Fig. \ref{fig:sensitivity_curves}b, a 1.5-kt OBD with an angular resolution of $\sigma_\theta = 20^\circ$ over a 20-year measurement has a median significance exceeding $2\sigma$ for rejecting the chemically homogeneous mantle hypothesis when the LLVP U enrichment factor exceeds $E^\mathrm{U}=10$.
This enrichment level corresponds to an LLVP HPE budget of approximately 25\% or more of the total mantle HPE budget.
The lack of statistical exclusion at lower $E^\mathrm{U}$ values is expected, because a small $E^\mathrm{U}$ inherently represents a highly homogenized mantle, making the $E^\mathrm{U}=1$ null hypothesis a reflection of the actual physical state rather than an observational limitation.

Taken together, these results demonstrate that directional geoneutrino detection can provide quantitative sensitivity to HPE enrichment in LLVPs and can exclude a chemically homogeneous mantle model when the enrichment is large.
Note that the significance shown here is not necessarily optimized. Future studies employing a more optimized angular binning scheme, including the azimuthal information, could further improve the statistical sensitivity.
In addition, this study adopts a fixed LLVP geometry and does not explicitly explore the uncertainties associated with its shape, spatial extent, or mass fraction.
The potential of energy spectral information to provide complementary constraints on deep-mantle HPE heterogeneity is also left for future investigation.
Incorporating these additional dimensions of information would provide a more comprehensive assessment of the capability of directional geoneutrino measurements to characterize deep-mantle compositional heterogeneity.

\section{Conclusions}
\label{sec:conclusion}

In this study, we investigate the sensitivity of directional geoneutrino detection to heat-producing element (HPE) enrichment within the deep mantle, using the tomographically inferred LLVP geometry as a specific test case.
In contrast to conventional rate-only measurements, directional information of geoneutrinos can improve the discrimination between an HPE-enriched LLVP scenario and a homogeneous mantle under the assumed detector response.
We find that an Ocean Bottom Detector (OBD) can provide a particularly favorable geological platform to determine whether LLVPs are chemical anomalies.
By leveraging the thin and HPE-depleted oceanic crust to suppress local backgrounds, the OBD achieves a mantle-to-total geoneutrino signal ratio exceeding 70\%.
This favorable signal-to-background environment provides a suitable setting for directly probing the HPE distribution within the mantle.

Our analysis shows that the sensitivity to HPE enrichment within the adopted LLVP geometry is governed by the interplay between detector's angular resolution, experimental time, background levels, and absolute magnitude of HPE enrichment.
Specifically, angular information can improve the discrimination between an HPE-enriched LLVP scenario and a homogeneous mantle scenario, demonstrating the importance of angular resolution, in addition to detector location, overburden, background levels, detector size, and experimental duration, for future directional geoneutrino experiments.
Further studies incorporating more detailed LLVP geometries, azimuthal information, energy spectra, and comparisons with alternative models of deep-mantle HPE distribution (e.g., HPE-enriched D$''$-layer model), will be important for assessing the full potential of directional geoneutrino measurements to constrain deep-mantle compositional heterogeneity.

\printcredits

\section*{Declaration of Competing Interest}
The authors declare that they have no known competing financial interests or personal relationships that could have appeared to influence the work reported in this paper.

\section*{Acknowledgments}
The authors would like to thank Shuai Ouyang and Prof. Eiji Ohtani for their valuable discussions, as well as all members of the OBD Consortium for their continuous support and cooperation. We also thank Prof. Vincent Salters and an anonymous reviewer for their constructive comments and suggestions, which helped improve this manuscript. This work was supported by JSPS KAKENHI Grant No. 24K00653 and the Sasakawa Scientific Research Grant No. 2026-6050 from The Japan Science Society. WFM gratefully acknowledges National Science Foundation (NSF) support (EAR2050374).

\appendix
\section{Statistical Framework and Profile Likelihood Analysis}
\label{app:statistics}
Given the low-statistics nature of geoneutrino detection, we utilize the binned Poisson deviance to evaluate the goodness-of-fit between the expected model $E_i$ and the mock observational data $O_i$.
To account for the uncertainties in the backgrounds, we consider Gaussian constraints on the nuisance parameters, yielding the complete profiled test statistic
\begin{equation}
    \chi^2_{\mathrm{profile}} =
    \min_{\bm{\epsilon}}
    \left[
    2 \sum_{i} \left( E_i - O_i + O_i \ln \left(\frac{O_i}{E_i}\right) \right)
    + \sum_{j} \left( \frac{\epsilon_j}{\sigma_j} \right)^2
    \right].
\end{equation}
Here, $\epsilon_j$ represents the deviation of the $j$-th background source from its nominal value, and $\sigma_j$ is its predefined relative uncertainty.

To quantify the median experimental sensitivity without the stochastic bias of random fluctuations, we employ the Asimov dataset introduced in \cite{cowan2011asymptotic} as our representative mock observation $O_i$.
In this limit, the observed counts in each bin are set exactly equal to the underlying model expectation, which ensures that the resulting best-fit parameters align with the prescribed values and the estimated sensitivities represent the median of the statistical ensemble.

The global minimum of $\chi^2_{\mathrm{profile}}$ defines the best-fit parameters. Confidence intervals and confidence regions are determined from
\begin{equation}
\Delta\chi^2 = \chi^2_{\mathrm{profile}} - \chi^2_{\mathrm{profile,min}},
\end{equation}
which asymptotically follows a $\chi^2$ distribution according to Wilks' theorem \citep{wilks1938}. For one parameter of interest, the confidence intervals corresponding to coverage probabilities of 68.3\%, 95.4\%, and 99.7\% are defined by $\Delta\chi^2 = 1.0$, $4.0$, and $9.0$, respectively. For two parameters of interest, the corresponding confidence regions are defined by $\Delta\chi^2 = 2.30$, $6.18$, and $11.83$. All nuisance parameters are profiled simultaneously.

\section{Monte Carlo Inference in Low-Statistics Regimes}
\label{app:beyond_Wilks}

The asymptotic approximation provided by Wilks' theorem \citep{wilks1938} can become inaccurate in the low-statistics regime, which is relevant for early-stage geoneutrino observations, analyses with fine angular binning, and inference near parameter boundaries.
To evaluate statistical significance without relying on the asymptotic $\chi^2$ approximation, we employ a frequentist Monte Carlo (MC) approach to construct the empirical distribution of the test statistic under the null hypothesis.

We generate $10^5$ mock experiments under the null hypothesis ($H_0$) and $10^4$ mock experiments under the alternative hypothesis ($H_1$).
For each mock experiment, the HPE abundances, crustal geoneutrino flux, and non-geoneutrino backgrounds are randomly sampled according to their respective uncertainties, and the observed event counts in each angular bin are generated by Poisson sampling.
The median experimental sensitivity is evaluated from the median $\Delta\chi^2$ obtained from the $H_1$ ensemble, with the central 68\% interval of the $H_1$ distribution used to quantify the corresponding spread.
The statistical significance is calibrated against the empirical $\Delta\chi^2$ distribution obtained from the $H_0$ ensemble.
The $p$-value is defined as the probability under $H_0$ of obtaining a test statistic greater than or equal to the observed value,
\begin{equation}
p = P_{H_0}\left(\Delta\chi^2 \geq \Delta\chi^2_{\rm obs}\right).
\end{equation}
When a sufficient number of $H_0$ MC realizations populate the relevant tail, the $p$-value is obtained directly from the empirical survival function.

However, direct MC counting becomes inefficient for very small $p$-values, since the number of realizations required to adequately sample the extreme tail increases rapidly.
To address this, we define a threshold $u$ as the 95th percentile of the empirical $H_0$ distribution, above which the tail is modeled using an exponential survival function.
The $p$-value for an observed value above this threshold is then obtained by anchoring the exponential tail to the empirically determined probability of exceeding $u$:
\begin{equation}
p =
P_{H_0}(\Delta\chi^2>u)
\exp\left[-b(\Delta\chi^2_{\rm obs}-u)\right],
\qquad
\Delta\chi^2_{\rm obs}>u,
\end{equation}
where $b$ is the exponential decay constant determined from the upper tail of the $H_0$ MC distribution.
The resulting $p$-values are converted into equivalent one-sided Gaussian significances according to
\begin{equation}
Z = \Phi^{-1}(1-p),
\end{equation}
where $\Phi^{-1}$ denotes the inverse cumulative distribution function of the standard normal distribution.

\bibliographystyle{cas-model2-names}
\setlength{\bibsep}{0ex}
\bibliography{cas-refs}

@article{mcdonough2026earth,
  title={Earth's composition: {O}rigin, energy budget, and insights from geoneutrinos},
  author={McDonough, William F.},
  journal={Geochim. Cosmochim. Acta},
  year={2026},
  volume={422},
  pages={9--34},
  doi={10.1016/j.gca.2025.12.060}
}

@article{ferrick2023defining,
title = {Defining {E}arth's elusive thermal budget in the presence of a hidden reservoir},
journal = {Earth Planet. Sci. Lett.},
volume = {601},
pages = {117893},
year = {2023},
doi = {10.1016/j.epsl.2022.117893},
author = {Amy L. Ferrick and Jun Korenaga},
}

@article{lyubetskaya2007a,
  title={Chemical composition of {E}arth's primitive mantle and its variance: 1. {M}ethod and results},
  author={Lyubetskaya, Tanya and Korenaga, Jun},
  journal={J. Geophys. Res. Solid Earth},
  volume={112},
  number={B3},
  year={2007},
  pages={B03211},
  doi = {10.1029/2005JB004223}
}

@article{lyubetskaya2007b,
  title={Chemical composition of {E}arth's primitive mantle and its variance: 2. {I}mplications for global geodynamics},
  author={Lyubetskaya, Tanya and Korenaga, Jun},
  journal={J. Geophys. Res. Solid Earth},
  volume={112},
  number={B3},
  year={2007},
  pages={B03212},
  doi={10.1029/2005JB004224}
}

@article{salters2004composition,
  title={Composition of the depleted mantle},
  author={Salters, Vincent JM and Stracke, Andreas},
  journal={Geochem. Geophys. Geosyst.},
  volume={5},
  pages={Q05B07},
  year={2004},
  doi={10.1029/2003GC000597}
}

@article{workman2005major,
  title={Major and trace element composition of the depleted {MORB} mantle ({DMM})},
  author={Workman, Rhea K and Hart, Stanley R},
  journal={Earth Planet. Sci. Lett.},
  volume={231},
  number={1-2},
  pages={53--72},
  year={2005},
  doi = {10.1016/j.epsl.2004.12.005}
}

@article{raghavan1998measuring,
  title={Measuring the global radioactivity in the {E}arth by multidetector antineutrino spectroscopy},
  author={Raghavan, R Si and Schoenert, S and Enomoto, S and Shirai, J and Suekane, F and Suzuki, A},
  journal={Phys. Rev. Lett.},
  volume={80},
  number={3},
  pages={635},
  year={1998},
  doi={10.1103/PhysRevLett.80.635}
}

@article{araki2005experimental,
  title={Experimental investigation of geologically produced antineutrinos with {KamLAND}},
  author={Araki, T and Enomoto, S and Furuno, K and others},
  collaboration={KamLAND Collaboration},
  journal={Nature},
  volume={436},
  number={7050},
  pages={499--503},
  year={2005},
  doi={10.1038/nature03980}
}

@article{eder1966terrestrial,
  title={Terrestrial neutrinos},
  author={Eder, Gernot},
  journal={Nucl. Phys.},
  volume={78},
  number={3},
  pages={657--662},
  year={1966},
  doi={10.1016/0029-5582(66)90903-5}
}

@article{garnero2016continent,
  title={Continent-sized anomalous zones with low seismic velocity at the base of {E}arth's mantle},
  author={Garnero, Edward J and McNamara, Allen K and Shim, Sang-Heon},
  journal={Nat. Geosci.},
  volume={9},
  number={7},
  pages={481--489},
  year={2016},
  doi={10.1038/ngeo2733}
}

@article{kamland2022geonu,
  title={Abundances of uranium and thorium elements in {E}arth estimated by geoneutrino spectroscopy},
  author={Abe, S and Asami, S and Eizuka, M and others},
  collaboration={KamLAND Collaboration},
  journal={Geophys. Res. Lett.},
  volume={49},
  number={16},
  pages={e2022GL099566},
  year={2022},
  doi={10.1029/2022GL099566}
}

@article{borexino2020geonu,
  title={Comprehensive geoneutrino analysis with {B}orexino},
  author={Agostini, M and Altenm{\"u}ller, K and Appel, S and others},
  collaboration={Borexino Collaboration},
  journal={Phys. Rev. D},
  volume={101},
  number={1},
  pages={012009},
  year={2020},
  doi={10.1103/PhysRevD.101.012009}
}

@article{sno2026geonu,
  title = {Measurement of reactor antineutrino oscillations with 1.46 kilotonne-years of data at {SNO}+},
  author={Abreu, M and Allega, A and Anderson, MR and others},
  collaboration = {SNO+ Collaboration},
  journal = {Phys. Rev. Lett.},
  year = {2026},
  doi = {10.1103/mv5z-pfb7},
  volume = {137},
  pages = {101807}
}

@article{juno2025geonu,
  title={Measurement of reactor neutrino oscillation with the first {JUNO} data},
  author={Abusleme, Angel and Adam, Thomas and Adamowicz, Kai and others},
  collaboration={JUNO Collaboration},
  journal={Nature},
  volume={654},
  pages={343–348},
  year={2026},
  doi={10.1038/s41586-026-10538-z}
}

@article{ANDES2012potential,
  title={Potential of a neutrino detector in the {ANDES} underground laboratory for geophysics and astrophysics of neutrinos},
  author={Machado, Pedro Accioly Nogueira and M{\"u}hlbeier, T and Nunokawa, Hiroshi and Funchal, R Zukanovich},
  journal={Phys. Rev. D},
  volume={86},
  number={12},
  pages={125001},
  year={2012},
  doi={10.1103/PhysRevD.86.125001}
}

@article{baksan2017whitepaper,
  title={Large-volume detector at the {Baksan Neutrino Observatory} for studies of natural neutrino fluxes for purposes of geo-and astrophysics},
  author={Barabanov, IR and Bezrukov, LB and Veresnikova, AV and others},
  journal={Phys. At. Nucl.},
  volume={80},
  number={3},
  pages={446--454},
  year={2017},
  doi={10.1134/S1063778817030036}
}

@article{nueye2026whitepaper,
  title={The $\nu${EYE} {Neutrino Telescope}: {C}onceptual Design Report.},
  author={Chen, Shaomin and Ahn, YangHwan and Franco, Davide and others},
  eprint={2601.12569},
  archivePrefix={arXiv},
  year={2026}
}

@article{obd2026whitepaper,
  title={Deep-Ocean Application-Specific Neutrino Experiment.},
  author={Araki, Takumi and Chauhan, Simran and Choi, Lyla and others},
  eprint={2606.20780},
  archivePrefix={arXiv},
  year={2026}
}

@article{tanaka20146li,
  title={{\textsuperscript{6}Li}-loaded directionally sensitive anti-neutrino detector for possible geo-neutrinographic imaging applications},
  author={Tanaka, H. K. M. and Watanabe, H.},
  journal={Sci. Rep.},
  volume={4},
  number={1},
  pages={4708},
  year={2014},
  doi={10.1038/srep04708}
}

@article{safdi2015directional,
  title={Directional antineutrino detection},
  author={Safdi, Benjamin R. and Suerfu, Burkhant},
  journal={Phys. Rev. Lett.},
  volume={114},
  number={7},
  pages={071802},
  year={2015},
  doi={10.1103/PhysRevLett.114.071802}
}

@article{seo2020labls,
  title = {Search for reactor neutrino directionality using a {LAB}-based {G}d-loaded liquid scintillation detector},
  author={Seo, Jun Hu and Atif, Zohaib and Shin, Chang Dong and Joo, Kyung Kwang},
  journal = {Nucl. Instrum. Methods Phys. Res. A},
  volume = {969},
  pages = {164001},
  year = {2020},
  doi = {10.1016/j.nima.2020.164001}
}

@article{sutanto2021sandd,
  title={{SANDD}: {A} directional antineutrino detector with segmented 6{L}i-doped pulse-shape-sensitive plastic scintillator},
  author={Sutanto, F and Classen, TM and Dazeley, SA and Duvall, MJ and Jovanovic, I and Li, VA and Mabe, AN and Reedy, ETE and Wu, T},
  journal={Nucl. Instrum. Methods Phys. Res. A},
  volume={1006},
  pages={165409},
  year={2021},
  doi={10.1016/j.nima.2021.165409}
}

@article{duvall2024directional,
  title={Directional response of several geometries for reactor-neutrino detectors},
  author={Duvall, Mark J and Crow, Brian C and Dornfest, Max AA and Learned, John G and Bergevin, Marc F and Dazeley, Steven A and Li, Viacheslav A},
  journal={Phys. Rev. Appl.},
  volume={22},
  number={5},
  pages={054030},
  year={2024},
  doi={10.1103/PhysRevApplied.22.054030}
}

@article{yepez2026algorithm,
  title={Algorithm to extract direction in 2{D} discrete distributions and a continuous {F}robenius norm},
  author={Yepez, Jeffrey G and Seligman, Jackson D and Dornfest, Max AA and Crow, Brian C and Learned, John G and Li, Viacheslav A},
  journal={AIP Adv.},
  volume={16},
  number={2},
  year={2026},
  doi={10.1063/5.0315079}
}

@article{crow2026enhancing,
  title = {Enhancing angular sensitivity of segmented antineutrino detectors for reactor monitoring},
  author={Crow, Brian C and Dornfest, Max AA and Learned, John G and Seligman, Jackson D and Sibert, Nathan S and Yepez, Jeffrey G and Li, Viacheslav A},
  journal = {Phys. Rev. Appl.},
  volume = {26},
  issue = {2},
  pages = {024019},
  year = {2026},
  doi = {10.1103/jrmk-r81s},
}

@misc{sun2026potassium,
  title = {Potassium-40 geoneutrinos detection and the {E}arth's large-scale structures imaging by directional geoneutrino detection.},
  author={Sun, Haozhe and Wang, Zhe and Chen, Shaomin},
  year={2026},
  eprint={2604.00808},
  archivePrefix={arXiv},
}

@article{shimizu2007directional,
  title={Directional measurement of anti-neutrinos},
  author={Shimizu, I.},
  journal={Nucl. Phys. B, Proc. Suppl.},
  volume={168},
  pages={147--149},
  year={2007},
  doi={10.1016/j.nuclphysbps.2007.02.071}
}

@article{leyton2017exploring,
  title={Exploring the hidden interior of the {E}arth with directional neutrino measurements},
  author={Leyton, Michael and Dye, Stephen and Monroe, Jocelyn},
  journal={Nat. Commun.},
  volume={8},
  number={1},
  pages={15989},
  year={2017},
  doi={10.1038/ncomms15989}
}

@article{fiorentini2012mantle,
  title={Mantle geoneutrinos in {K}am{LAND} and {B}orexino},
  author={Fiorentini, Giovanni and Fogli, Gian Luigi and Lisi, Eligio and Mantovani, Fabio and Rotunno, Anna Maria},
  journal={Phys. Rev. D},
  volume={86},
  number={3},
  pages={033004},
  year={2012},
  doi={10.1103/PhysRevD.86.033004}
}

@article{sammon2022quantifying,
  title={Quantifying {E}arth's radiogenic heat budget},
  author={Sammon, Laura G. and McDonough, William F.},
  journal={Earth Planet. Sci. Lett.},
  volume={593},
  pages={117684},
  year={2022},
  doi={10.1016/j.epsl.2022.117684}
}

@article{vsramek2016revealing,
  title={Revealing the {E}arth's mantle from the tallest mountains using the {J}inping {N}eutrino {E}xperiment},
  author={{\v{S}}r{\'a}mek, Ond{\v{r}}ej and Roskovec, Bed{\v{r}}ich and Wipperfurth, Scott A and Xi, Yufei and McDonough, William F},
  journal={Sci. Rep.},
  volume={6},
  number={1},
  pages={33034},
  year={2016},
  doi={10.1038/srep33034}
}

@article{xu2026proceedings,
	title = {Towards imaging {Earth}'s large-scale structures by directional geoneutrino detection with {Ocean} {Bottom} {Detector}},
	author={Xu, Zhihao and Araki, Takumi and Chauhan, Simran and others},
    collaboration={OBD Consortium},
    doi = "10.22323/1.511.0315",
    journal = "PoS Proc. Sci.",
    year = 2026,
    volume = 511,
    pages = "315"
}

@article{mcdonough2003compositional,
  title={Compositional model for the {E}arth's core},
  author={McDonough, William F.},
  journal={Treatise on Geochem.},
  volume={2},
  pages={568},
  year={2003},
  doi={10.1016/B0-08-043751-6/02015-6}
}

@article{wipperfurth2018earth,
  title={Earth's chondritic {Th/U}: {N}egligible fractionation during accretion, core formation, and crust--mantle differentiation},
  author={Wipperfurth, Scott A and Guo, Meng and {\v{S}}r{\'a}mek, Ond{\v{r}}ej and McDonough, William F},
  journal={Earth Planet. Sci. Lett.},
  volume={498},
  pages={196--202},
  year={2018},
  doi={10.1016/j.epsl.2018.06.029}
}

@article{sakai2021obd,
  title={Study of {O}cean {B}ottom {D}etector for observation of geo-neutrino from the mantle},
  author={Sakai, T and Inoue, K and Watanabe, H and others},
  journal={J. Phys. Conf. Ser.},
  volume={2156},
  number={1},
  pages={012144},
  year={2021},
  doi={10.1088/1742-6596/2156/1/012144}
}

@article{huang2013reference,
  title={A reference {E}arth model for the heat-producing elements and associated geoneutrino flux},
  author={Huang, Yu and Chubakov, Viacheslav and Mantovani, Fabio and Rudnick, Roberta L and McDonough, William F},
  journal={Geochem. Geophys. Geosyst.},
  volume={14},
  number={6},
  pages={2003--2029},
  year={2013},
  doi={10.1002/ggge.20129}
}

@article{wipperfurth2020reference,
  title={Reference models for lithospheric geoneutrino signal},
  author={Wipperfurth, Scott A and {\v{S}}r{\'a}mek, Ond{\v{r}}ej and McDonough, William F},
  journal={J. Geophys. Res. Solid Earth},
  volume={125},
  number={2},
  pages={e2019JB018433},
  year={2020},
  doi={10.1029/2019JB018433}
}

@inproceedings{watanabe2023obd,
  title={Ocean {B}ottom {D}etector: frontier of technology for understanding the mantle by geoneutrinos},
  author={Watanabe, Hiroko and Abe, Natsue and Araki, Eiichiro and others},
  booktitle={2023 IEEE Underwater Technology (UT)},
  pages={1--5},
  year={2023},
  doi={10.1109/UT49729.2023.10103417}
}

@article{arevalo2010chemical,
  title={Chemical variations and regional diversity observed in {MORB}},
  author={Arevalo Jr., Ricardo and McDonough, William F.},
  journal={Chem. Geol.},
  volume={271},
  number={1-2},
  pages={70--85},
  year={2010},
  doi={10.1016/j.chemgeo.2009.12.013}
}

@article{arevalo2013simplified,
  title={Simplified mantle architecture and distribution of radiogenic power},
  author={Arevalo Jr, Ricardo and McDonough, William F and Stracke, Andreas and Willbold, Matthias and Ireland, Thomas J and Walker, Richard J},
  journal={Geochem. Geophys. Geosyst.},
  volume={14},
  number={7},
  pages={2265--2285},
  year={2013},
  doi={10.1002/ggge.20152}
}

@article{sramek2013geophysical,
  title={Geophysical and geochemical constraints on geoneutrino fluxes from {E}arth's mantle},
  author={{\v{S}}r{\'a}mek, Ond{\v{r}}ej and McDonough, William F and Kite, Edwin S and Leki{\'c}, Vedran and Dye, Stephen T and Zhong, Shijie},
  journal={Earth Planet. Sci. Lett.},
  volume={361},
  pages={356--366},
  year={2013},
  doi={10.1016/j.epsl.2012.11.001}
}

@article{cottaar2016morphology,
  title={Morphology of seismically slow lower-mantle structures},
  author={Cottaar, Sanne and Lekic, Vedran},
  journal={Geophys. J. Int.},
  volume={207},
  number={2},
  pages={1122--1136},
  year={2016},
  doi={10.1093/gji/ggw324}
}

@phdthesis{enomoto2005phd,
  author       = {Enomoto, Sanshiro},
  title        = {Neutrino geophysics and observation of geo-neutrinos at {KamLAND}},
  school       = {Tohoku University},
  year         = {2005},
  address      = {Sendai, Japan},
  type         = {{Ph.D. Thesis}}
}

@article{strumia2003ibd,
  title={Precise quasielastic neutrino/nucleon cross-section},
  author={Strumia, Alessandro and Vissani, Francesco},
  journal={Phys. Lett. B},
  volume={564},
  number={1-2},
  pages={42--54},
  year={2003},
  doi={10.1016/S0370-2693(03)00616-6}
}

@article{dziewonski1981prem,
  title={Preliminary reference {E}arth model},
  author={Dziewonski, Adam M. and Anderson, Don L.},
  journal={Phys. Earth Planet. Inter.},
  volume={25},
  number={4},
  pages={297--356},
  year={1981},
  doi={10.1016/0031-9201(81)90046-7}
}

@article{mcdonough1995composition,
  title={The composition of the {E}arth},
  author={McDonough, William F. and Sun, S-S},
  journal={Chem. Geol.},
  volume={120},
  number={3-4},
  pages={223--253},
  year={1995},
  doi={10.1016/0009-2541(94)00140-4}
}

@article{cowan2011asymptotic,
  title={Asymptotic formulae for likelihood-based tests of new physics},
  author={Cowan, Glen and Cranmer, Kyle and Gross, Eilam and Vitells, Ofer},
  journal={Eur. Phys. J. C},
  volume={71},
  number={2},
  pages={1554},
  year={2011},
  doi={10.1140/epjc/s10052-011-1554-0}
}

@article{liquido2026geonu,
  title={Probing {E}arth's missing potassium using the antimatter signature of geoneutrinos},
  author={Cabrera, A. and Chen, M. and Mantovani, F. and others},
  collaboration={LiquidO Collaboration},
  journal={Commun. Phys.},
  year={2026},
  volume={9},
  pages={95},
  doi={10.1038/s42005-026-02518-6}
}

@article{theia2022geonu,
  title={Geo-and reactor antineutrino sensitivity at {THEIA}},
  author={Zsoldos, Stephane and Bagdasarian, Zara and Orebi Gann, Gabriel D and Barna, Andrew and Dye, Stephen},
  journal={Eur. Phys. J. C},
  volume={82},
  number={12},
  pages={1151},
  year={2022},
  doi={10.1140/epjc/s10052-022-11106-1}
}

@article{roskovec2018testing,
  title={Testing a proposed "second continent" beneath eastern {C}hina using geoneutrino measurements.},
  author={Roskovec, Bedrich and {\v{S}}r{\'a}mek, Ond{\v{r}}ej and McDonough, William F},
  eprint={1810.10914},
  archivePrefix={arXiv},
  year={2018}
}

@article{PDF2024cfk,
    author = "Navas, S. and Amsler, C. and Gutscheand, T. and others",
    collaboration = "Particle Data Group",
    title = "Review of particle physics",
    doi = "10.1103/PhysRevD.110.030001",
    journal = "Phys. Rev. D",
    volume = "110",
    number = "3",
    pages = "030001",
    year = "2024"
}

@article{mohr2025codata,
  title={{CODATA} recommended values of the fundamental physical constants: 2022},
  author={Mohr, Peter J and Newell, DB and Taylor, BN and others},
  journal={J. Phys. Chem. Ref. Data},
  volume={54},
  number={3},
  pages={033105},
  year={2025},
  doi={10.1063/5.0279860}
}

@misc{dye2015global,
      title={Global antineutrino modeling for a web application.}, 
      author={Stephen T. Dye and Andrew Barna},
      year={2015},
      eprint={1510.05633},
      archivePrefix={arXiv},
}

@article{giunti2022reactor,
      title={Reactor antineutrino anomaly in light of recent flux model refinements},
      author={Giunti, C. and Li, YF and Ternes, CA and Xin, Z},
      journal={Phys. Lett. B},
      volume={829},
      pages={137054},
      year={2022},
      doi={10.1016/j.physletb.2022.137054}
}

@article{vogel1999angular,
  title={Angular distribution of neutron inverse beta decay, $\overline{\nu_e} + p \rightarrow e^+ + n$},
  author={Vogel, P and Beacom, John F},
  journal={Phys. Rev. D},
  volume={60},
  number={5},
  pages={053003},
  year={1999},
  doi={10.1103/PhysRevD.60.053003}
}

@article{wang2020hunting,
  title={Hunting potassium geoneutrinos with liquid scintillator Cherenkov neutrino detectors},
  author={Wang, Zhe and Chen, Shaomin},
  journal={Chin. Phys. C},
  volume={44},
  number={3},
  pages={033001},
  year={2020},
  doi={10.1088/1674-1137/44/3/033001}
}

@article{white2014,
  title={Composition of the oceanic crust},
  author={White, WM and Klein, EM},
  journal={Treatise on Geochem.},
  volume={4},
  pages={457--496},
  year={2014},
  doi={10.1016/B978-0-08-095975-7.00315-6}
}

@article{palme2014,
    title={Cosmochemical Estimates of Mantle Composition},
    author={Palme, H and O’Neill, H},
    journal={Treatise on Geochem.},
    volume={3},
    pages={1--39},
    year={2014},
	doi={10.1016/B978-0-08-095975-7.00201-1},
}

@article{rudnick2014,
    title = {Composition of the continental crust},
    journal={Treatise on Geochem.},
    volume={4},
    pages={1--51},
    year={2014},
    doi = {10.1016/B978-0-08-095975-7.00301-6},
    author = {R.L. Rudnick and S. Gao}
}

@article{plank2014,
    title = {The chemical composition of subducting sediments},
    journal={Treatise on Geochem.},
    volume={4},
    pages={607--629},
    year={2014},
    doi = {10.1016/B978-0-08-095975-7.00319-3},
    author = {T. Plank}
}

@article{mooney2023ecm,
  title={Earth {C}rustal {M}odel 1 ({ECM}1): {A} 1 x 1 global seismic and density model},
  author={Mooney, Walter D and Barrera-Lopez, C. and Suárez, MG and Castelblanco, MA},
  journal={Earth-Sci. Rev.},
  volume={243},
  pages={104493},
  year={2023},
  doi={10.1016/j.earscirev.2023.104493}
}

@article{kamland2015solarnu,
  title={$^{7}${B}e solar neutrino measurement with {K}am{LAND}},
  author={Gando, A. and Gando, Y. and Hanakago, H. and others},
  collaboration={KamLAND Collaboration},
  journal={Phys. Rev. C},
  volume={92},
  pages={055808},
  year={2015},
  doi={10.1103/PhysRevC.92.055808}
}

@article{mcdonough2020radiogenic,
  title={Radiogenic power and geoneutrino luminosity of the {E}arth and other terrestrial bodies through time},
  author={McDonough, William F and {\v{S}}r{\'a}mek, Ond{\v{r}}ej and Wipperfurth, SA},
  journal={Geochem. Geophys. Geosyst.},
  volume={21},
  number={7},
  pages={e2019GC008865},
  year={2020},
  doi={10.1029/2019GC008865}
}

@article{fukuda1998evidence,
  title={Evidence for oscillation of atmospheric neutrinos},
  author={Fukuda, Yoshiyuki and Hayakawa, T. and Ichihara, E. and others},
  collaboration={Super-Kamiokande Collaboration},
  journal={Phys. Rev. Lett.},
  volume={81},
  number={8},
  pages={1562},
  year={1998},
  doi={10.1103/PhysRevLett.81.1562}
}

@article{wilks1938,
  title={The large-sample distribution of the likelihood ratio for testing composite hypotheses},
  author={Wilks, Samuel S},
  journal={Ann. Math. Statist.},
  volume={9},
  number={1},
  pages={60--62},
  year={1938},
  url={https://www.jstor.org/stable/2957648}
}

@article{fields2006imaging,
  title={Imaging the {E}arth’s interior: the angular distribution of terrestrial neutrinos},
  author={Fields, Brian D and Hochmuth, Kathrin A},
  journal={Earth Moon Planets},
  volume={99},
  number={1},
  pages={155--181},
  year={2006},
  doi={10.1007/s11038-006-9132-4}
}

@article{ritsema2011s40rts,
  title={{S40RTS}: a degree-40 shear-velocity model for the mantle from new {Rayleigh} wave dispersion, teleseismic traveltime and normal-mode splitting function measurements},
  author={Ritsema, J. and Deuss, A. and van Heijst, H. J. and Woodhouse, J. H.},
  journal={Geophys. J. Int.},
  volume={184},
  number={3},
  pages={1223--1236},
  year={2011},
  doi={10.1111/j.1365-246X.2010.04884.x}
}

@article{auer2014savani,
  title={Savani: {A} variable resolution whole-mantle model of anisotropic shear velocity variations based on multiple data sets},
  author={Auer, L. and Boschi, L. and Becker, T. W. and Nissen-Meyer, T. and Giardini, D.},
  journal={J. Geophys. Res. Solid Earth},
  volume={119},
  number={4},
  pages={3006--3034},
  year={2014},
  doi={10.1002/2013JB010773}
}

@article{french2014whole,
  title={Whole-mantle radially anisotropic shear velocity structure from spectral-element waveform tomography},
  author={French, SW and Romanowicz, Barbara A},
  journal={Geophys. J. Int.},
  volume={199},
  number={3},
  pages={1303--1327},
  year={2014},
  doi={10.1093/gji/ggu334}
}

@article{tesoniero2015hydration,
  title={Hydration of marginal basins and compositional variations within the continental lithospheric mantle inferred from a new global model of shear and compressional velocity},
  author = {Tesoniero, Andrea and Auer, Ludwig and Boschi, Lapo and Cammarano, Fabio},
  journal={J. Geophys. Res. Solid Earth},
  volume={120},
  number={11},
  pages={7789--7813},
  year={2015},
  doi={10.1002/2015JB012026}
}

@article{moulik2016relationships,
  title={The relationships between large-scale variations in shear velocity, density, and compressional velocity in the {E}arth's mantle},
  author={Moulik, P and Ekstr{\"o}m, G},
  journal={J. Geophys. Res. Solid Earth},
  volume={121},
  number={4},
  pages={2737--2771},
  year={2016},
  doi={10.1002/2015JB012679}
}



\end{document}